\documentclass[english,aps,prl,showpacs,superscriptaddress,floatfix, notitlepage,reprint,unicode=true,colorlinks=true,citecolor=Blue,linkcolor=RubineRed,urlcolor=Blue]{revtex4-2}
\usepackage[T1]{fontenc}
\usepackage[utf8]{inputenc}
\usepackage{color}
\usepackage{babel}
\usepackage{mathrsfs}
\usepackage{amsmath}
\usepackage{amsthm}
\usepackage{amssymb}
\usepackage{graphicx}
\usepackage[unicode=true]
 {hyperref}

\makeatletter
\usepackage{braket}
\usepackage{txfonts} 
\usepackage{graphicx}
\usepackage[usenames,dvipsnames]{xcolor}
\date{\today}

\makeatother

\begin{document}
\title{Universal Shot-Noise Limit for Quantum Metrology with Local Hamiltonians}
\author{Hai-Long Shi}
\address{Innovation Academy for Precision Measurement Science and Technology,
Chinese Academy of Sciences, Wuhan 430071, China}
\affiliation{QSTAR and INO-CNR, Largo Enrico Fermi 2, 50125 Firenze, Italy}
\address{Hefei National Laboratory, Hefei 230088, China}
\author{Xi-Wen Guan}
\address{Innovation Academy for Precision Measurement Science and Technology,
Chinese Academy of Sciences, Wuhan 430071, China}
\address{Hefei National Laboratory, Hefei 230088, China}
\address{Department of Fundamental and Theoretical Physics, Research School
of Physics, Australian National University, Canberra ACT 0200, Australia}
\author{Jing Yang\href{https://orcid.org/0000-0002-3588-0832}{\includegraphics[scale=0.05]{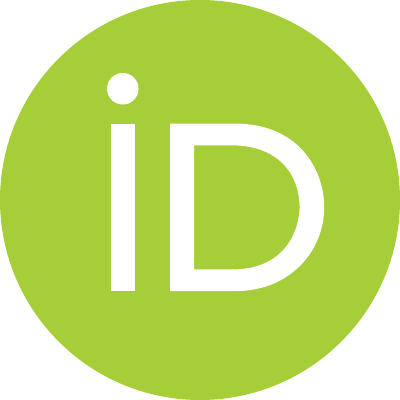}}}
\email{jing.yang@su.se}
\address{Nordita, KTH Royal Institute of Technology and Stockholm University,
Hannes Alfv\'ens vag 12, 10691 Stockholm, Sweden}
\begin{abstract}
Quantum many-body interactions can induce quantum entanglement among
particles, rendering them valuable resources for quantum-enhanced sensing. 
In this work, we establish a link between the bound on the growth of the quantum Fisher information and the Lieb-Robinson bound, which characterizes the operator
growth in locally interacting quantum many-body systems. We 
show that for initial separable states, despite the use of local many-body interactions, the precision cannot surpass the shot noise limit at all
times. 
This conclusion also holds for an initial state that is the nondegenerate
ground state of a local and gapped Hamiltonian. 
These findings strongly hint that when one can only prepare separable initial states, nonlocal
and long-range interactions are essential resources for surpassing the shot
noise limit.
This observation is confirmed through numerical analysis on the long-range
Ising model.
Our results bridge the field of many-body quantum sensing
and operator growth in many-body quantum systems and open the possibility
to investigate the interplay between quantum sensing and control,
many-body physics and information scrambling.
\end{abstract}
\maketitle
\textit{Introduction}.\textit{---}Quantum entanglement is a valuable
resource in quantum information processing. 
In quantum metrology,
quantum Fisher information (QFI)\,\citep{helstrom1976quantum,holevo2011probabilistic,liu2019quantum,paris2009quantum,braun2018quantumenhanced},
quantifying the precision of the sensing parameter, scales linearly
with the number for uncorrelated probes, known as
the shot noise limit (SNL),
which also appears in sensing with classical resources. 
Quantum entanglement can achieve the Heisenberg limit (HL), featuring quadratic scaling, or even surpass it to reach the super-HL. 
Entanglement manifests its efficacy in two primary ways: during state preparation\,\citep{giovannetti2006quantum,giovannetti2011advances,demkowicz-dobrzanski2014usingentanglement,pezze2009entanglement,toth2012multipartite,hyllus2012fisherinformation}
or during signal sensing via the many-body interactions among
individual sensors\,\citep{boixo2007generalized,roy2008exponentially,yang2022,yang2022superheisenberg,beau2017nonlinear},
which is the main essence of many-body quantum metrology. 
Recently,
the subject matter has gained renewed interest. However, the existing
protocols of dynamic sensing require to prepare the initial state in the highly entangled
Greenberger--Horne--Zeilinger (GHZ)-like states, whose preparation
is very challenging and time-consuming. An effective strategy to address
this issue involves combining the protocols of quantum state preparation
and quantum metrology, see, e.g., Refs.\,\citep{chu2023strongquantum,dooley2016quantum,hayes2018makingthe},
where an entangled initial state is prepared before the sensing process.
Nevertheless, evaluating the time required to prepare a highly entangled state from separable ones, while considering restrictions imposed by accessible Hamiltonians, proves to be extremely challenging\,\citep{yang2022minimumtime,bukov2019geometric,carlini2006timeoptimal}.
On the other hand, the protocols of quantum critical sensing either necessitate an initial state that near the vicinity of quantum criticality\,\citep{zanardi2008quantum, Gu2008Fidelity, rams2018atthe,frerot2018quantum,garbe2020critical,montenegro2021global,sahoo2023localization,10.21468/SciPostPhys.13.4.077,CAROLLO20201,Carollo_2019,leonforte2019uhlmann} or involve  critical quantum dynamics\,\citep{chu2021dynamic, wan2023quantum}, which is time consuming to reach due to critical
slow down. 
As such, the time required for initial state preparation in quantum critical sensing is also largely ignored\,\citep{gietka2021adiabatic}.

\begin{figure}
\begin{centering}
\includegraphics[scale=0.35]{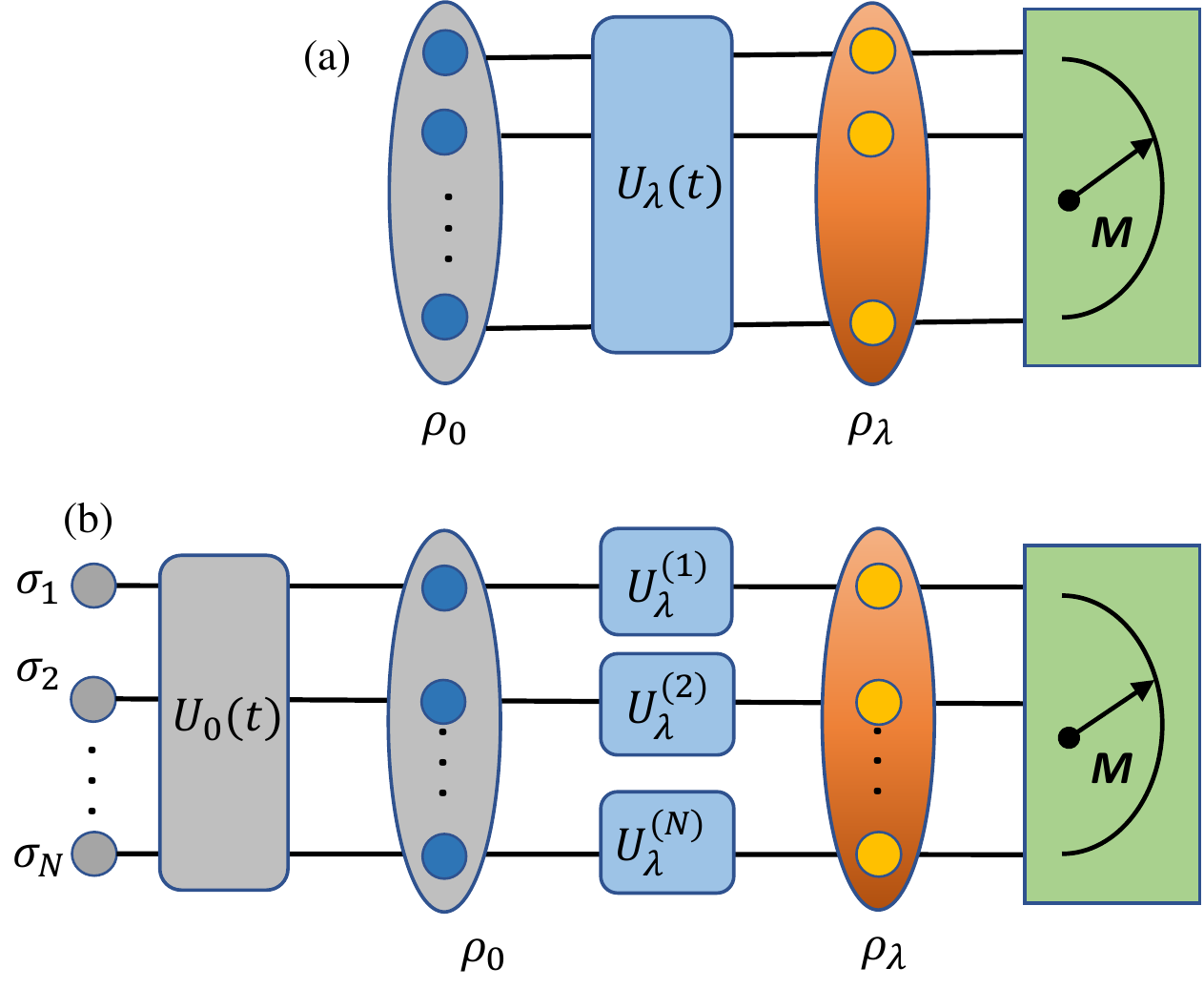}
\par\end{centering}
\caption{\label{fig:Comparison}Comparison between our protocol (a) with the
protocol in Ref.~\citep{chu2023strongquantum}\,(b). In our protocol\,(a), the information of the estimation parameter is encoded into the
many-body quantum states through the many-body dynamics $U_{\lambda}(t)=e^{-\text{i}(\lambda\sum_{i}h_{X_{i}}+H_{1})t}$
while in Ref.\,\citep{chu2023strongquantum}, the encoding dynamics
given by $U_{\lambda}=e^{-\text{i}\lambda\sum_{i}h_{X_{i}}}$ with
$X_{j}=\{j\}$. In our protocol, the initial state is chosen to be
either a separable state or the nondegenerate ground states of a
gapped and local Hamiltonian while in Ref.\,\citep{chu2023strongquantum}
the initial state is prepared through the many-body dynamics $U_{0}(t)$.}
\end{figure}

To circumvent the overhead of quantum state preparation, in this work,  we propose to prepare the probes
or sensors initially in a separable state, which can be prepared with
the current experimentally feasible technology\,\citep{livingston2022experimental,Blumoff2022,an2022highfidelity}.
In our protocol depicted in Fig.\,\ref{fig:Comparison}(a), entanglement
emerges during the signal sensing process due to the interactions
in the many-body sensing Hamiltonian. 
This contrasts sharply with
the protocol in Ref.\,\citep{chu2023strongquantum} illustrated in
Fig.\,\ref{fig:Comparison}(b), where the entangled initial state
is explicitly prepared through the time evolution driven by a locally
interacting preparation Hamiltonian, while the sensing Hamiltonian
is noninteracting. 

It is well known in the literature that for separable initial states and a noninteracting sensing Hamiltonian, the precision is limited by the SNL\,\citep{giovannetti2006quantum, giovannetti2011advances, demkowicz-dobrzanski2014usingentanglement}. 
In our protocol, due to the many-body interactions, the state can become entangled after the sensing process.
This prompts the central  question:  whether many-body interactions can break the SNL. 
This question is also intimately related to recent studies on operator growth and quantum chaos in quantum many-body systems\,\citep{parker2019auniversal, balasubramanian2022quantum, noh2021operator, rabinovici2021operator, barbon2019onthe}.

To answer this question, we derive a universal bound governing the growth
of QFI over time, which can characterize
the role of quantum entanglement in information scrambling, operator
growth, and quantum chaos.
We apply our bound to dynamic quantum sensing
protocols with time-independent many-body Hamiltonians as shown in
Fig.\,\ref{fig:Comparison}(a) and estimate the bound using the celebrated
Lieb-Robinson bound\,\citep{lieb1972thefinite,bravyi2006liebrobinson,nachtergaele2009liebrobinson,hastings2004locality}
for quantum many-body systems with local interactions. We find that
it is impossible to surpass the SNL with local interactions. This 
observation holds not only for separable initial states but also
extends to cases where the initial state is the nondegenerate ground state of
a locally gapped Hamiltonian---a state feasible for experimental preparation through cooling processes.
Therefore, if only separable states are accessible in
experiments, nonlocal or long-range interactions are essential to
beat the SNL and bring real quantum advantage in many-body quantum
metrology. We exemplify our findings in magnetometry with the short-range
transverse-field Ising (TFI) model, the chaotic Ising (CI) model,
and the long-range Ising (LRI) model.

\textit{Universal bound on the growth of the QFI. ---}We consider
the following sensing Hamiltonian 
\begin{flalign}
H_{\lambda}(t)=H_{0\lambda}(t)+H_{1}(t),\label{eq:generic-H}
\end{flalign}
where $H_{0\lambda}(t)$ is a simple Hamiltonian encoding the estimation parameter $\lambda$, and $H_{1}(t)$
involves interactions among sensors induced by either intrinsic interactions
or external coherent controls. In the formal case, $H_{1}$ is usually
time independent, while in the later case, $H_{1}(t)$ becomes time dependent.
The generator for sensing $\lambda$\,\citep{boixo2007generalized,pang2017optimal}
is given by
\begin{flalign}
G(t)=\int_{0}^{t}\left[\partial_{\lambda}H_{\lambda}(\tau)\right]^{(\text{H})}d\tau,\label{eq:G}
\end{flalign}
where an operator in the Heisenberg picture is defined as $\mathcal{O}^{(\text{H})}(t)\!=\!U^{\dagger}(t)\mathcal{O}^{(\text{S})}(t)U(t)$.
The QFI is determined by the variance of $G(t)$ over the initial
state $\ket{\psi_{0}}$, i.e., 
\begin{flalign}
I(t)=\text{4Var}[G(t)]{}_{\ket{\psi_{0}}}.\label{eq:QFI}
\end{flalign}
Optimal control theory has been proposed to simultaneously optimize
the initial state $\ket{\psi_{0}}$ and $H_{1}(t)$, resulting in
a bound $I(t)\!\leq\!4\left(\int_{0}^{t}\|\partial_{\lambda}H_{\lambda}^{(\text{S})}(\tau)\|d\tau\right)^{2}$\,\citep{pang2017optimal,yang2017quantum,yang2022,yuan2015optimal}.
Here, the seminorm $\|\cdot\|$ denotes the spectrum width
of an operator, i.e., the difference between its maximum eigenvalue
and minimum eigenvalue.

By taking the derivative of Eq.\,(\ref{eq:QFI}) and applying the Cauchy-Schwarz inequality,  we derive a universal bound\,\citep{SM,PhysRevLett.131.160801} that characterizes the growth of QFI:
\begin{flalign}
\frac{d\sqrt{I(t)}}{dt}\leq\Gamma(t)\equiv2\sqrt{\text{Var}\left(\left[\partial_{\lambda}H_{\lambda}(t)\right]^{(\text{H})}\right){}_{\ket{\psi_{0}}}}.\label{eq:Universal-bound}
\end{flalign}
The saturation condition is provided
in the Supplemental Material (SM)\,\citep{SM}. Alternatively, one can
rewrite 
\begin{flalign}
\Gamma(t)=2\sqrt{\text{Var}\left(\partial_{\lambda}H_{\lambda}^{\text{(S)}}(t)\right){}_{\ket{\psi(t)}}},\label{eq:Gamma-psi-t}
\end{flalign}
where $\ket{\psi(t)}\!=\!U(t)\ket{\psi_{0}}$.

It is worth noting that Eq.\,(\ref{eq:Universal-bound}) universally holds  for all initial states, including  time-independent and driven quantum systems.
 $\Gamma(t)$
depends on the control Hamiltonian $H_{1}(t)$ and the initial state
$\ket{\psi_{0}}$. Optimizing $\Gamma(t)$ over all possible unitary
dynamics and initial states yields $\Gamma(t)\!\leq\!2\|\partial_{\lambda}H_{\lambda}^{\text{(S)}}(t)\|$.
By combining this bound with $I(t)\!\leq\!\left(\int_{0}^{t}\Gamma(\tau)d\tau\right)^{2}$,
which can be obtained by integrating both sides of Eq.\,(\ref{eq:Universal-bound}),
one immediately reproduces the bound given in previous works\,\citep{yuan2015optimal,pang2017optimal,yang2017quantum,yang2022}.
Compared to these studies, our bound\,(\ref{eq:Universal-bound})
provides a feasible approach to study the scaling behavior of the
QFI when the initial state $\ket{\psi_{0}}$ is limited to a specific set of states.

\textit{SNL for short-range local interactions.---}
We will show the close connection between our bound\,(\ref{eq:Universal-bound}), depicting QFI growth, and the Lieb-Robinson bound, which characterizes operator complexity
in quantum many-body with short-range local interactions. 
We consider time-independent Hamiltonians as follows:
\begin{flalign}
H_{\lambda}=\lambda\sum_{i=1}^{N}h_{X_{i}}+H_{1},\label{eq:time-indept-H}
\end{flalign}
where $h_{X_{i}}$ is supported on the set $X_{i}$ with cardinality
$|X_{i}|\!=\!R$ and diameter $\text{diam}(X_{i})\!=\!\max_{k,\,l\in X_{j}}|k-l|$.
$H_{1}$ denotes the interactions. 
We require
$H_{\lambda}$ to contain only local and short-range interactions, imposing
that $\text{diam}(X_{j})$ is independent of $N$\,\citep{hastings2004locality,bravyi2006liebrobinson,zeng2019quantum}
and $h_{X_{j}}$ is a local operator. Equation\,(\ref{eq:time-indept-H})
represents the model used in magnetometry, where $\lambda$ represents the
magnetic field\,\citep{degen2017quantum}. 

According to Eq.\,(\ref{eq:time-indept-H}), the bound\,(\ref{eq:Universal-bound})
can be reformulated in relation to dynamic correlation matrices of local operators as
\begin{flalign}
\Gamma(t)=2\sqrt{\sum_{jk}\text{Cov}[h_{X_{j}}^{(\text{H})}(t)h_{X_{k}}^{(\text{H})}(t)]{}_{\ket{\psi_{0}}}},\label{eq:gamma-cov}
\end{flalign}
where $\text{Cov}[AB]\big|_{\ket{\psi_{0}}}\!\equiv\!\langle\{A,\,B\}\rangle/2\!-\!\langle A\rangle\langle B\rangle$
and 
\begin{flalign}
h_{X_{i}}^{(\text{H})}(t)\!\equiv\! e^{\text{i}H_{\lambda}t}h_{X_{i}}e^{-\text{i}H_{\lambda}t}.\label{eq:ht}
\end{flalign}
In this case, we observe that
$\Gamma(t)\!\leq\!2N,$ implying $I(t)\!\leq\!4N^{2}t^{2}$\,\citep{yuan2015optimal,pang2017optimal}.
If $H_{1}$ commutes with $\sum_{i=1}^{N}h_{X_{i}}$, then such an HL can be saturated only when using GHZ-like entangled initial states\,\citep{yang2022,yang2022superheisenberg,boixo2007generalized}.
However, preparing such states experimentally is challenging.
Conversely, if $H_{1}$ does not commute with $\sum_{i=1}^{N}h_{X_{i}}$, then entanglement maybe generated by signal
sensing from separable initial states.
So, for separable initial states, what precisely is the tight bound that limits the precision? 
Is it possible to surpass the SNL using many-body interactions?

We emphasize that the Lieb-Robinson bound\,\citep{nachtergaele2009liebrobinson,lieb1972thefinite,bravyi2006liebrobinson,hastings2004locality} imposes a strong restriction on the scaling of QFI for local Hamiltonians.
Specifically, if the sensing Hamiltonian\,(\ref{eq:time-indept-H})
only contains local or short-range interactions, the static correlation
$\text{Cov}[h_{X_{j}}h_{X_{k}}]\big|_{\ket{\psi_{0}}}$ between two
disjoint local operators $h_{X_{j}}$ and $h_{X_{k}}$ decays exponentially,
provided the initial state $\ket{\psi_{0}}$ is separable or the nondegenerate
ground state of some local and gapped Hamiltonians.
In this case, the dynamic
correlation function also decays exponentially, 
\begin{flalign}
\big|\text{Cov}[h_{X_{j}}^{(\text{H})}(t)h_{X_{k}}^{(\text{H})}(t)]{}_{\ket{\psi_{0}}}\big|\leq\mathcal{C}\exp(-[d(X_{j},\,X_{k})-v_{\text{LR}}t]/\xi),\label{eq:correlation-LR-bound}
\end{flalign}
where $\mathcal{C}$ and $\xi$ are constants that solely depend on
the topology of the sites, $d(X_{j},\,X_{k})$ is the distance between
$X_{j}$ and $X_{k}$, and $v_{\text{LR}}$ is the celebrated Lieb-Robinson
velocity. 
Substituting Eq.\,(\ref{eq:correlation-LR-bound}) into Eq.\,(\ref{eq:gamma-cov}), we rigorously show
that the scaling of $\Gamma(t)$ is lower bounded by $\sqrt{N}$. The crucial observation here is that upon factoring out the time-dependent term $\exp(v_{\text{LR}}t/\xi)$ , thanks to the exponential decay of dynamic correlation functions, only initially overlapping local operators will contribute to the scaling of $\Gamma(t)$. This results in\,\citep{SM}
\begin{flalign}
\Gamma(t)\leq2\gamma(t)\sqrt{N},\label{eq:Gamma-LR-bound}
\end{flalign}
where $\gamma(t)$ is only a function of time and independent of $N$.
It remains finite as long as $t$ is finite and behaves as $e^{v_{\text{LR}}t/\xi}$
as $t\!\to\!\infty$.

Equation\,(\ref{eq:Gamma-LR-bound}) is the main
result of this work. Clearly, for finite but fixed times, the QFI is limited by the SNL. 
On the
other hand, at sufficiently long times, for time-independent systems,
one can show that $I(t)/t^{2}$ is independent of time\,\citep{SM,pang2014quantum}
and is only a function of $N$. 
Moreover,
the timescale to reach this regime corresponds to the case where
$t$ is much larger than the inverse of the minimum energy gap for
the system. In this regime, when $N$ is large, $I(t)\sim t^{2}N^{\alpha}$.
Since Eq.\,(\ref{eq:Gamma-LR-bound}) is valid for all times and all
$N$, combined with Eq.\,(\ref{eq:Universal-bound}) we conclude $\alpha\leq1$.
Therefore, in local short-range models where operator growth is constrained
by the Lieb-Robinson bound, the SNL cannot be surpassed. 

Nevertheless, for the same initial state, the many-body interaction $H_1$ may increase the prefactor of the QFI compared to the noninteracting case, though it is not always the case. 
For example, if the initial state is prepared in a state slightly deviating from the ground state of $\sum_{i=1}^{N}h_{X_{i}}$, then the QFI for the noninteracting case, being the fluctuations of $G(t)$ over the initial state, grows very slowly as time evolves. 
Meanwhile, if a many-body interaction that does not commute with the noninteracting Hamiltonian is added, the noncommutativity introduces significant fluctuations of $G(t)$, which can lead to a QFI  significantly larger than the noninteracting case and thus enhance the prefactor of QFI.

\textit{The spread of the generator of the metrological bound. ---
}
The manifestation of the SNL in locally interacting systems and for separable
initial states can also be  understood from the perspective of operator growth. 
Despite $h_{X_{i}}^{\text{(H})}(t)$ spreads over the lattice, the
metrological generator $\left[\partial_{\lambda}H_{\lambda}(t)\right]^{(\text{H})}$,
being a sum of these nonlocal operators, may still be reformulated as a sum of local operators, thus keeping the precision limited to the SNL. 
A trivial example is
when $H_{1}$ commutes with $\sum_{i}h_{X_{i}}$ while $H_{\lambda}$
does not commute with each individual $h_{X_{i}}$, in which case
$\Gamma(t)$ remains at the SNL.

Generally, we assume $\left[\partial_{\lambda}H_{\lambda}(t)\right]^{\text{(}\text{H})}$
can be expanded in terms of two-body basis operators
\begin{flalign}
\left[\partial_{\lambda}H_{\lambda}(t)\right]^{(\text{H})}=\sum_{i=1}^{N}\sum_{j\ge i}^{N}\sum_{\alpha}\eta_{ij}^{\alpha}\mathcal{O}_{ij}^{\alpha}\label{eq:operator-decomposition}
\end{flalign}
where we have suppressed the time dependence for simplicity and for
spin systems $\mathcal{O}_{ij}^{\alpha}$ is a basis spin operator,
such as $\sigma_{i}^{x}\sigma_{j}^{y}$ while for fermionic systems
$\mathcal{O}_{ij}^{\alpha}$ is a Hermitian basis fermionic operator,
such as $c_{i}^{\dagger}c_{j}+\text{H.C.}$or $\text{i}(c_{i}^{\dagger}c_{j}-\text{H.C.})$.
We emphasize that the number of different types of operators
indexed by $\alpha$ is finite and does not scale with $N$. If the
initial state is separable and $\left[\partial_{\lambda}H_{\lambda}(t)\right]^{(\text{H})}$
is a sum of fast-decaying long-range two-body interactions, i.e.,
\begin{flalign}
\lim_{k\to\infty}\lim_{N\to\infty}\sum_{i\leq k}\sum_{j\ge k}|\eta_{ij}^{\alpha}|=\lim_{k\to\infty}\int_{1}^{k}dx\int_{k}^{\infty}dy|\eta_{xy}^{\alpha}|<\infty,\label{eq:eta-decay}
\end{flalign}
then the SNL cannot be surpassed by using the bound\,(\ref{eq:Universal-bound})\,\citep{SM}. 
It follows from Eq.\,(\ref{eq:operator-decomposition}) that 
$\left[\partial_{\lambda}H_{\lambda}(t)\right]^{(\text{H})}\!=\!\sum_{i=1}^{N}\tilde{\mathscr{O}}_{i}$,
where $\tilde{\mathscr{O}}_{i}\!\equiv\!\frac{1}{2}\sum_{\alpha}\left(\sum_{j\ge i}^{N}\eta_{ij}^{\alpha}\mathcal{O}_{ij}^{\alpha}+\sum_{j\leq i}^{N}\eta_{ji}^{\alpha}\mathcal{O}_{ji}^{\alpha}\right)$.
The condition\,(\ref{eq:eta-decay}) ensures that $\tilde{\mathscr{O}}_{i}$
behaves effectively as a local
operator and can be different
from $h_{X_{i}}^{(\text{H})}(t)$, which is generically nonlocal.
Essentially, the locality of components of the metrological generator leads to the SNL for separable initial states.
We will further elaborate this observation using the TFI model. 

\textit{SNL in the TFI model.---}We consider the integrable TFI chain 
\begin{flalign}
H_{\lambda}^{{\rm TFI}}=-\left(J\sum_{i=1}^{N}\sigma_{i}^{x}\cdot\sigma_{i+1}^{x}+\lambda\sum_{i=1}^{N}\sigma_{i}^{z}\right),\label{eq:integrable-Ising-chain}
\end{flalign}
with the periodic boundary condition $\sigma_{1}^{z}\!=\!\sigma_{N+1}^{z}$,
and $J,\,\lambda\!>\!0$. In the thermodynamic limit $N\!\to\!\infty$,
when $J\!\gg\!\lambda$ the ground state is ferromagnetic and degenerate,
represented by $|++\cdots+\rangle$ or $|--\cdots-\rangle$, while
for $J\!\ll\!\lambda$ the ground state is paramagnetic $|\uparrow\uparrow\cdots\uparrow\rangle$. 

For any initial separable state, \textcolor{black}{Eq.\,(\ref{eq:Universal-bound})}
predicts that the QFI cannot surpass the SNL. On the other hand, \textcolor{black}{this
model can be exactly solved by mapping it to a free fermion model}\,\citep{LIEB1961407,PFEUTY197079,carollo2018symmetric} and therefore one can compute $\left[\partial_{\lambda}H_{\lambda}(t)\right]^{(\text{H})}$
explicitly. 
In our SM\,\citep{SM}, we demonstrate that the metrological generator $\left[\partial_{\lambda}H_{\lambda}(t)\right]^{(\text{H})}$ in this case explicitly follows the structure of Eq.\,(\ref{eq:operator-decomposition}) with
four types of fermionic operators: $\mathcal{O}_{ij}^{(1)}\!=\!(c_{i}^{\dagger}c_{j}+{\rm H.C.})$,
$\mathcal{O}_{ij}^{(2)}\!=\!(c_{i}c_{j}^{\dagger}+{\rm H.C.)}$, $\mathcal{O}_{ij}^{(3)}\!=\!(c_{i}^{\dagger}c_{j}^{\dagger}+{\rm H.C.}),$
and $\mathcal{O}_{ij}^{(4)}\!=\!\text{i}(c_{i}^{\dagger}c_{j}^{\dagger}-{\rm H.C.})$.
The expression for the $\eta$ functions characterizing the weights
of these operators spreading from the $i$th site to the $j$th
site can be found in the SM\,\citep{SM}. In the thermodynamic limit, $\eta_{ij}^{\alpha}$
behaves like $p^{j-i}$ for $j\ge i$, where $p\!=\!J/\lambda$ for $J\!<\!\lambda$,
$p\!=\!\lambda/J$ for $\lambda\!<\!J$, and $p\!=\!0$ for $J\!=\!\lambda$\,\citep{SM}.
The power-law decay of $\eta$ functions indicates that the evolved
operator remains extremely local, as shown in Fig.\,\ref{fig:TFI-chain}(b),
ensuring the condition\,(\ref{eq:eta-decay}), i.e., 
\begin{flalign}
\text{\ensuremath{\int_{1}^{k}dx\int_{k}^{\infty}dy|\eta_{xy}^{\alpha}|\sim\text{\ensuremath{\frac{1-p^{k-1}}{(\ln p)}}}<\infty}},
\end{flalign}
as $k\!\to\!\infty.$ Therefore, the locality of the evolved operator suggests
that QFI beyond the SNL cannot be achieved by initial separable probe
states in this integrable TFL model. Figure \,\ref{fig:TFI-chain}(a)
characterizes the diffusion of the correlators, suggesting that the
numerical choices of $t=0.5$ and $t=5\times10^{4}$ can be considered
as the timescales for the part and full spread of local operators, respectively.
Figures\,\ref{fig:TFI-chain}(c) and\,\ref{fig:TFI-chain}(d) numerically verify that only the SNL
can be achieved for the different initial separable spin coherent
states parameterized by $\ket{\psi_{0}}=\bigotimes_{i=1}^{N}[\cos(\theta/2)\ket{\uparrow}_{i}+\sin(\theta/2)e^{{\rm i\phi}}\ket{\downarrow}_{i}]$.

Furthermore, if we consider the initial state as the ground state of the
TFI model with known values of parameters $\lambda_{*}$ and $J$,
achievable through cooling processes, then the asymptotic behavior of the QFI
with respect to the unknown parameter $\lambda$ under the the Hamiltonian\,(\ref{eq:integrable-Ising-chain})
is 
\begin{flalign}
\lim_{t,N\to\infty}\frac{I(t)}{Nt^{2}}=f(\lambda,J,\lambda_{*})\sim O(1),\label{eq:QFI-quench}
\end{flalign}
where the function $f$ is $N$ independent\,\citep{SM1}, confirming the claim that
only the SNL can be achieved even with the ground state of local and gapped
Hamiltonians. Taking $\lambda_{*}\!\to\!+\infty$, where the ground state
becomes the spin coherent state with $\theta\!=\!\phi\!=\!0$, we find 
\begin{flalign}
\lim_{t,N\to\infty}\frac{I(t)}{Nt^{2}}=\frac{J^{2}(4\lambda^{2}-3J^{2})}{\lambda^{4}}\label{eq:ANA}
\end{flalign}
for $J\!<\!\lambda$, which is also verified in Fig.\,\ref{fig:TFI-chain}(d), and $\lim_{t,N\to\infty}I(t)/(Nt^{2})\!=\!1$ for $J\!\ge\!\lambda$.
We observe that the prefactor of the QFI, both for the ground state and for other separable states [as depicted in Figs.\,\ref{fig:TFI-chain}(c)-\,\ref{fig:TFI-chain}(f)], does not exhibit a significant difference in order of magnitude, i.e., <10, when compared to the optimal noninteraction scheme involving separable states, where $I(t)/(Nt^2)\!=\!\|\sigma_i^z\|^2\!=\!4.$

\begin{figure}
\begin{centering}
\includegraphics[width=1\columnwidth]{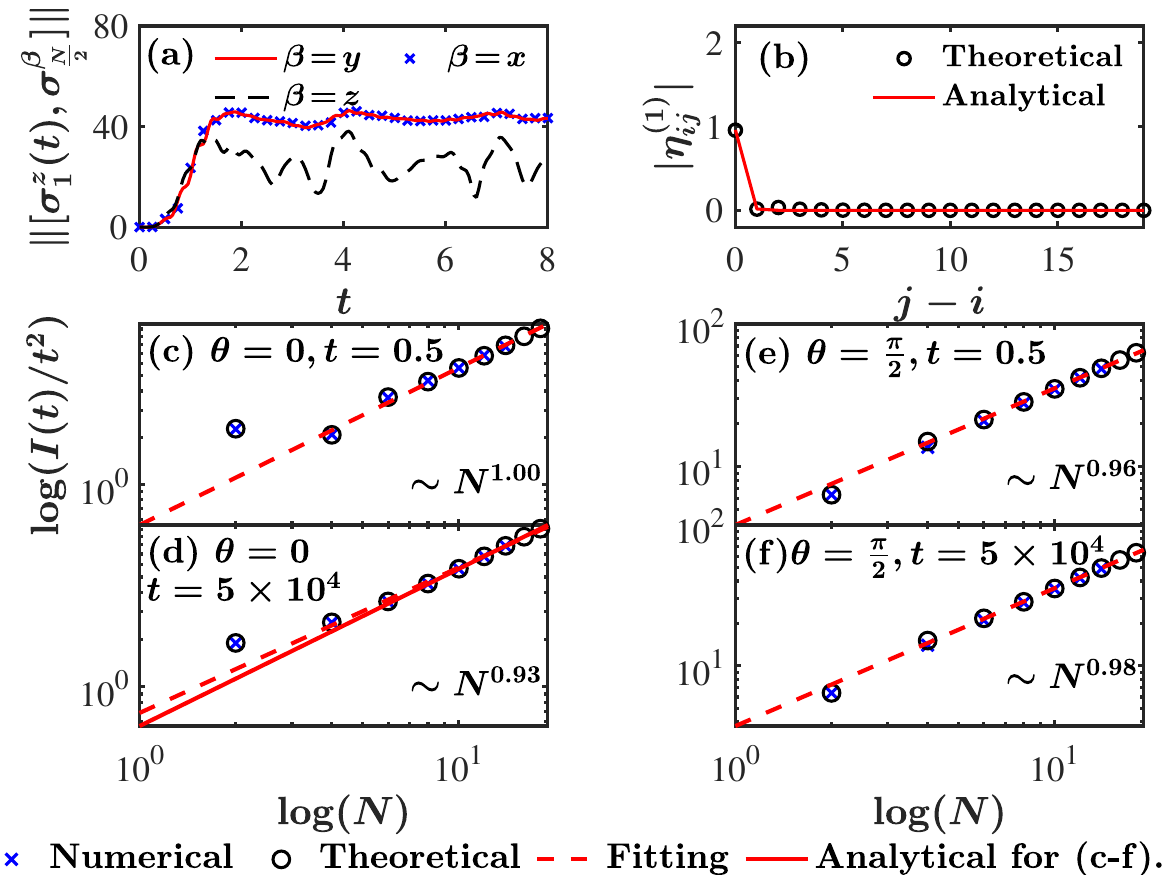}
\par\end{centering}
\caption{\label{fig:TFI-chain}(a)\,Numerical calculation of the operator diffusion
in the TFI chain with $N\!=\!10$. (b)\,Coefficient $|\eta_{ij}^{(1)}|$
characterizing the decay of the two-body interactions. (c)-(f)\,Scaling
of the QFI with respect to the number of spins at different times
for differential initial separable spin coherent states $\ket{\psi_{0}}\!=\!\bigotimes_{i=1}^{N}[\cos(\theta/2)\ket{\uparrow}_{i}+\sin(\theta/2)e^{{\rm i\phi}}\ket{\downarrow}_{i}]$.
Here numerical data are obtained by directly diagonalizing the Hamiltonian
of the TFI model, while theoretical data are derived using results
by mapping the TFI model to the free fermion model. The analytical
result refers to Eq.\,(\ref{eq:ANA}). Other parameters used for
the calculations are $J\!=\!2,$ $\lambda\!=\!5$, and $\phi\!=\!0$.}
\end{figure}

\textit{SNL in the chaotic Ising model.---}
Different from integrable
models, the operator complexity in chaotic models grows very rapidly\,\citep{parker2019auniversal,balasubramanian2022quantum,rabinovici2021operator,barbon2019onthe,noh2021operator}.
Nevertheless, according to Eqs.\,(\ref{eq:Universal-bound}) and (\ref{eq:Gamma-LR-bound}), even in locally chaotic models, the SNL cannot be surpassed by using separable
states. 
For instance, we consider the Ising model with both
transverse and longitudinal fields described by the following Hamiltonian:
\begin{flalign}
H_{\lambda}^{{\rm CI}}=-\sum_{i=1}^{N}(J\sigma_{i}^{x}\sigma_{i+1}^{x}+h\sigma_{i}^{x}+\lambda\sigma_{i}^{z}),
\end{flalign}
where open boundary conditions are adopted. Energy-level spacing statistics
indicate that this model is quantum chaotic for $J\!=\!h\!=\!\lambda$\,\citep{noh2021operator,karthik2007entanglement}.
Figures\,\ref{fig:chaoticIsing-LRI}(c)-\,\ref{fig:chaoticIsing-LRI}(f)\,verify the prediction by Eq.\,(\ref{eq:Gamma-LR-bound})
that separable states cannot surpass the SNL even in such chaotic
short-range systems. To surpass the SNL, we are thus motivated to
explore long-range models. The effect of quantum chaos in quantum
metrology has been studied in Ref.\,\citep{fiderer2018quantum} within
the context of kick top, which involves long-range interactions.
Here, we show
that quantum chaos plays no enhancement in the scaling of QFI in locally chaotic many-spin models.

\textit{Beyond the SNL with the LRI model.---}
As demonstrated before, breaking the SNL is solely feasible within long-range and nonlocal systems, which violates the Lieb-Robinson inspired bound\,(\ref{eq:Gamma-LR-bound}).
Thus, we consider the long-range Ising model with power-law decay,
\begin{flalign}
H_{\lambda}^{{\rm LRI}}=-\left(J\sum_{i<j}\frac{\sigma_{i}^{x}\sigma_{j}^{x}}{|i-j|^{\alpha}}+\lambda\sum_{i}\sigma_{i}^{z}\right),
\end{flalign}
which reduces to the TFI model as $\alpha\!\to\!\infty.$ For $\alpha\!=\!0$,
this model corresponds to the Lipkin-Meshkov-Glick model\,\citep{lipkin1965validity}.
In this long-range \textcolor{black}{model, the breakdown of exponential
decay in connected correlation function Eq.\,(\ref{eq:correlation-LR-bound})
will result in the failure of the bound presented in\,(\ref{eq:Gamma-LR-bound}).
Consequently, we expect that for small $\alpha$ where the long-range
interactions decay sufficiently slowly, it is possible to surpass the
SNL with separable initial states. As depicted in Figs.\,\ref{fig:chaoticIsing-LRI}(g) and\,\ref{fig:chaoticIsing-LRI}(h),
we have identified specific instances of this scenario.}

\begin{figure}
\begin{centering}
\includegraphics[width=1\columnwidth]{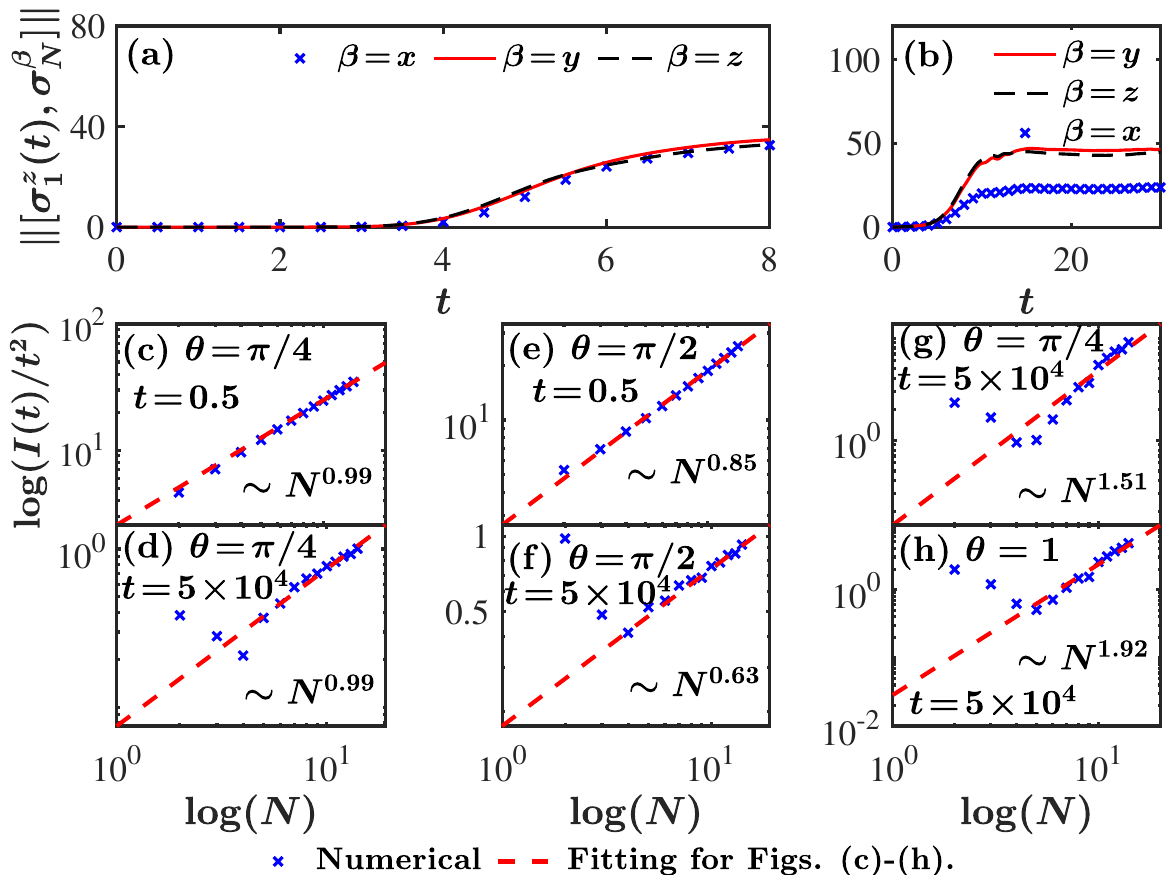}
\par\end{centering}
\caption{\label{fig:chaoticIsing-LRI}Numerical calculation of the operator
diffusion in (a)\,the CI model (b)\,the LRI model with $N\!=\!10$. The
scaling of the QFI with respect to the number of spins at different
times for differential initial separable spin coherent states $\ket{\psi_{0}}\!=\!\bigotimes_{i=1}^{N}[\cos(\theta/2)\ket{\uparrow}_{i}\!+\!\sin(\theta/2)e^{{\rm i\phi}}\ket{\downarrow}_{i}]$
in (c)-(f)\,the CI model and (g) and (h)\,the LRI model. Other parameters used
for the calculations are $J\!=\!\lambda\!=\!h\!=\!1,$ $\phi\!=\!0$ in the CI model,
and $J\!=\!1$, $\lambda\!=\!0.5,$ $\alpha\!=\!3$ in the LRI model.}
\end{figure}
\textit{Conclusion and outlook}.\textit{---}In conclusion, we have
derived a universal bound on the growth of the QFI under arbitrary
dynamics and initial states. We apply our bound to the case of separable
initial states or the nondegenerate ground state of a gapped and
local sensing Hamiltonian. 
We prove that with these particular initial states, the QFI cannot surpass the SNL, as we have explicitly
demonstrated with TFI and CI models. Our results give an important guideline for many-body sensing: either initial entanglement or long-range interactions are essential
resources to achieve quantum advantage in many-body quantum sensing, as demonstrated
in the LRI model. 
Our results shed light on various aspects of the interplay between many-body physics, quantum control theory, quantum chaos, operator growth, and information scrambling. We leave these studies for future exploration.

\emph{Notes added}.\textit{---}Recently, we noted that a bound similar to Eq.~\eqref{eq:Universal-bound} also appears in Ref.~\citep{PhysRevLett.131.160801} with the focus on non-Hermitian sensing.

\emph{Acknowledgement}.\textit{---}We thank Adolfo del Campo for
useful comments on the manuscript. It is a pleasure to acknowledge
feedback from Federcio Balducci, Wenchao Ge, Jiru Liu, Xingze Qiu, and Hong-Hao Tu. 
X.W.G. was supported by the NSFC key Grants No.~92365202, No.~12134015, No.~12121004, and partially supported by the Innovation Program
for Quantum Science and Technology~2021ZD0302000. 
H.L.S. was supported by the European Commission through the H2020 QuantERA ERA-NET Cofund in Quantum Technologies project “MENTA” and the Hefei National Laboratory.
J.Y. was funded by the Wallenberg Initiative on Networks and Quantum Information (WINQ)
and would like to thank Hui Zhai for the hospitality to host his 
visit at the Institute of Advanced Study in Tsinghua University, during
which this work was completed.

\let\oldaddcontentsline\addcontentsline     
\renewcommand{\addcontentsline}[3]{}         

\bibliographystyle{apsrev4-1}
\bibliography{Many-body-Metrology}

\begin{thebibliography}{65}
\bibitem [{SM1()}]{SM1}%
  \BibitemOpen
  \href@noop {} {}\bibinfo {note}
  {$f=(\lambda-\lambda{}_{*})(J^{2}-2\lambda\lambda_{*}+\lambda{}_{*}^{2})/(J^{2}-\lambda\lambda_{*})^{2}$
  for $0<\lambda,\lambda_{*}<J$, $f=J^{2}(\lambda-\lambda_{*})^{2}(\lambda
  J^{4}+2\lambda_{*}J^{4}-4\lambda^{2}\lambda_{*}J^{2}-3\lambda\lambda_{*}J^{2}+4\lambda^{3}\lambda_{*}{}^{2})/[\lambda_{*}{}^{2}\lambda^{3}(J^{2}-\lambda\lambda_{*})^{2}]$
  for $J<\lambda,\lambda_{*}$, $f=[2\lambda_{*}J^{2}-3\lambda
  J^{2}+\lambda(\lambda_{*}-2\lambda)^{2}]/\lambda^{3}$ for $
  0<\lambda_{*}<J<\lambda$, and
  $f=(J^{2}-2\lambda\lambda_{*}+\lambda_{*}{}^{2})/\lambda_{*}{}^{2}$ for
  $0<\lambda<J<\lambda_{*}$. Details can be found in the SM.}\BibitemShut
  {Stop}%
\end{thebibliography}

\clearpage\newpage\setcounter{equation}{0} \setcounter{section}{0}
\setcounter{subsection}{0} 
\global\long\def\theequation{S\arabic{equation}}%
\onecolumngrid \setcounter{enumiv}{0} 

\setcounter{equation}{0} \setcounter{section}{0} \setcounter{subsection}{0} \renewcommand{\theequation}{S\arabic{equation}} \onecolumngrid \setcounter{enumiv}{0}

\clearpage

\textbf{Supplemental Material for ``Universal shot-noise limit for quantum metrology with local Hamiltonians"}
\maketitle

\section{Abstract}
In this Supplemental Material, we present the saturation condition of the bound on the growth of the QFI, its connection to Lieb-Robinson bound, the asymptotic behavior of $G(t)/t$ at late times, the shot noise limit (SNL) for a metrological generator consisting of fast-decaying two-body operators, the analytic calculations of the metrological generator  and the quantum Fisher information (QFI) for the transverse-field Ising (TFI) model, respectively.

\section{\label{sec:Saturation-condition}Saturation condition of the bound
on the growth of the QFI}

We rewrite Eq.~(3) in the main text as 
\begin{equation}
I(t)=4\big\langle\left[G(t)-\langle G(t)\rangle\right]^{2}\big\rangle,
\end{equation}
where $\langle\bullet\rangle$ denotes the average on the initial
state $\ket{\psi_{0}}$. Taking derivatives on both sides, we find
\begin{align}
\dot{I}(t) & =4\big\langle\{\dot{G}(t)-\langle\dot{G}(t)\rangle,\,G(t)-\langle G(t)\rangle\}\big\rangle\nonumber \\
 & =\text{{\rm 8Re}\ensuremath{\langle}}[\dot{G}(t)-\langle\dot{G}(t)\rangle]\left[G(t)-\langle G(t)\rangle\right]\rangle\nonumber \\
 & \leq8|\langle[\dot{G}(t)-\langle\dot{G}(t)\rangle]\left[G(t)-\langle G(t)\rangle\right]\rangle|.
\end{align}
Using the Cauchy-Schwarz inequality, we find
\begin{equation}
\big|\big\langle[\dot{G}(t)-\langle\dot{G}(t)\rangle]\left[G(t)-\langle G(t)\rangle\right]\big\rangle\big|\leq\sqrt{\text{Var}[\dot{G}(t)]\text{Var}[G(t)]}.
\end{equation}
Thus, we find
\begin{equation}
\dot{I}(t)\leq4\sqrt{\text{Var}[\dot{G}(t)]I(t)},
\end{equation}
which is equivalent to the universal bound~(4)
in the main text. We define 
\begin{equation}
\ket{\Phi_{0}(t)}\equiv\left(G(t)-\langle G(t)\rangle\right)\ket{\psi_{0}},\label{eq:Phi0-def}
\end{equation}
with initial values 
\begin{equation}
\ket{\Phi_{0}(0)}=0,\quad\ket{\dot{\Phi}_{0}(0)}=\left[\partial_{\lambda}H_{\lambda}(0)-\langle\partial_{\lambda}H_{\lambda}(0)\rangle\right]\ket{\psi_{0}}.\label{eq:initial-condition}
\end{equation}
The inequality is saturated at some instant time $t$ if
\begin{equation}
\text{Im}\braket{\dot{\Phi_{0}}(t)\big|\Phi_{0}(t)}=0,\,\text{Re}\braket{\dot{\Phi_{0}}(t)\big|\Phi_{0}(t)}\ge0,\,,\,c(t)\ket{\dot{\Phi}_{0}(t)}=\ket{\Phi_{0}(t)},
\end{equation}
which is equivalent to 
\begin{equation}
c(t)\ket{\dot{\Phi}_{0}(t)}=\ket{\Phi_{0}(t)},\,c(t)\ge0.\label{eq:sat-diff}
\end{equation}
Now we require the inequality is satisfied from the initial time $t=0$
up to time $t$, we should integrate Eq.~(\ref{eq:sat-diff}). Note
that initial time $t=0$ may be not appropriate to be used as the reference
point for the integration since $\ket{\Phi_{0}(0)}=0$. Instead, we
can choose a reference point $t_{0}\in(0,\,t]$ such that $\ket{\Phi(t_{0})}\neq0$
and then obtain
\begin{equation}
\ket{\Phi_{0}(t)}=e^{\int_{t_{0}}^{t}\frac{d\tau}{c(\tau)}}\ket{\Phi_{0}(t_{0})}.\label{eq:generic-form}
\end{equation}
By further considering the requirement $c(\tau)\geq0$, we can conclude
that $\ket{\Phi_{0}(t)}$ should not change its direction as time
evolves, while exponentially increasing in its length. To satisfy
the initial condition~(\ref{eq:initial-condition}), we require that
\begin{equation}
\lim_{\epsilon\to0}e^{-\int_{\epsilon}^{t_{0}}\frac{d\tau}{c(\tau)}}=0,\,\lim_{\epsilon\to0}\frac{e^{-\int_{\epsilon}^{t_{0}}\frac{d\tau}{c(\tau)}}}{c(\epsilon)}=\alpha(t_{0}),\,\alpha(t_{0})\in\mathbb{R},\,\label{eq:IC-constraint}
\end{equation}
where 
\begin{equation}
\ket{\dot{\Phi}_{0}(0)}=\alpha(t_{0})\ket{\Phi(t_{0})}.
\end{equation}
The first condition of Eq.~(\ref{eq:IC-constraint}) implies that
$c(\epsilon)$ should vanish as $\epsilon\to0$. Thus we assume $c(\epsilon)\sim\epsilon^{\delta}$
as $\epsilon\to0$ with $\delta\ge1$. A straightforward analysis
show that the only possible case that is consistent with the second
equation of Eq.~(\ref{eq:IC-constraint}) is $\delta=1$. Therefore,
Eq.~(\ref{eq:generic-form}) may be rewritten as 
\begin{equation}
\ket{\Phi_{0}(t)}=\frac{e^{\int_{t_{0}}^{t}\frac{d\tau}{c(\tau)}}}{\alpha(t_{0})}\ket{\dot{\Phi}_{0}(0)}
\end{equation}
To summarize, Eq.~(\ref{eq:Phi0-def}) and Eq.~(\ref{eq:generic-form})
must be consistent. Taking derivatives on sides, we obtain the condition
for the saturation of the maximum growth rate of the QFI:
\begin{equation}
\left(U^{\dagger}(t)\partial_{\lambda}H_{\lambda}(t)U(t)-\langle U^{\dagger}(t)\partial_{\lambda}H_{\lambda}(t)U(t)\rangle\right)\ket{\psi_{0}}=\frac{e^{\int_{t_{0}}^{t}\frac{d\tau}{c(\tau)}}}{\alpha(t_{0})c(t)}\ket{\dot{\Phi}_{0}(0)},\,\forall t\label{eq:U-eq}
\end{equation}
which is obviously satisfied at $t=0$.

\section{\label{sec:LR-bound}Analysis of the growth bound of the QFI from
Lieb-Robinson bound}

It is straightforward to show that \textcolor{black}{
\begin{align}
\frac{1}{4}\Gamma^{2}(t) & =\sum_{j=1}^{N}\text{Var}[h_{X_{j}}^{(\text{H})}]{}_{\ket{\psi_{0}}}+\sum_{j,k=1,\,j\neq k}^{N}\text{Cov}[h_{X_{j}}^{(\text{H})}(t)h_{X_{k}}^{(\text{H})}(t)]{}_{\ket{\psi_{0}}}\nonumber \\
 & =\sum_{j=1}^{N}\text{Var}[h_{X_{j}}^{(\text{H})}(t)]{}_{\ket{\psi_{0}}}+\sum_{j=1}^{N}\sum_{k\in\mathcal{B}(j,\,R)}\text{Cov}[h_{X_{j}}^{(\text{H})}(t)h_{X_{k}}^{(\text{H})}(t)]{}_{\ket{\psi_{0}}}+\sum_{j=1}^{N}\sum_{k\notin\mathcal{B}(j,\,R)}\text{Cov}[h_{X_{j}}^{(\text{H})}(t)h_{X_{k}}^{(\text{H})}(t)]{}_{\ket{\psi_{0}}}\label{eq:Gam-sq}
\end{align}
where $R$ is the cardinality of the support $X_{j}$ and $\mathcal{B}(j,R)$
represents the closed neighborhood of site $j$ with a radius $R$.
Specifically, this implies that the intersection between the sets
$X_{j}$ and $X_{k}$ is empty if and only if the site $k$ does not
belong to the closed neighborhood $\mathcal{B}(j,R)$, i.e.,
\begin{equation}
X_{j}\cap X_{k}=\varnothing\,\iff\,k\notin\mathcal{B}(j,\,R).
\end{equation}
}The first term on the right-hand side of Eq.~(\ref{eq:Gam-sq})
describes the self-correlation of local operators. The second term
accounts for the correlation between pairs of local operators that
overlap from the initial time. The last term represents the
correlation among the pairs of local operators whose supports are
initially disjoint but may overlap later due to the Lieb-Robinson
type of diffusion.

Note that $\|h_{X_{j}}\|\sim R$, which does not scale with $N$.
As a result, we find that 
\begin{equation}
\text{Var}[h_{X_{j}}^{(\text{H})}(t)]{}_{\ket{\psi_{0}}}\text{\ensuremath{\lesssim R^{2},}}\label{eq:II-2}
\end{equation}
and 
\begin{equation}
\big|\sum_{k\in\mathcal{B}(j,\,R)}\text{Cov}[h_{X_{j}}^{(\text{H})}(t)h_{X_{k}}^{(\text{H})}(t)]{}_{\ket{\psi_{0}}}\big|\lesssim R^{3}.\label{eq:II-3}
\end{equation}
On the other hand, if $k\notin\mathcal{B}(j,\,R)$, then the set $X_{j}$
and $X_{k}$ are disjoint, allowing us to estimate $\text{Cov}[h_{X_{j}}^{(\text{H})}(t)h_{X_{k}}^{(\text{H})}(t)]{}_{\ket{\psi_{0}}}$
by using the Lieb-Robinson bound~(10).
Consequently, we find ~\citep{nachtergaele2009liebrobinson,lieb1972thefinite,bravyi2006liebrobinson,hastings2004locality},
\begin{equation}
\big|\sum_{k\notin\mathcal{B}(j,\,R)}\text{Cov}[h_{X_{j}}^{(\text{H})}(t)h_{X_{k}}^{(\text{H})}(t)]{}_{\ket{\psi_{0}}}\big|\lesssim R^{2}\sum_{k\notin\mathcal{B}(j,\,R)}\exp\left(-[d(X_{j},\,X_{k})-v_{\text{LR}}t]/\xi\right),\label{eq:II-4}
\end{equation}
which holds for initial separable states or the non- degenerate ground
state of local and gapped Hamiltonians. Here, $d(X_{j},X_{k})$ is
the distance connecting the set $X_{j}$ with the set $X_{k}$, $\xi$
is a constant that solely depends on the topology of the sites, and
$v_{\text{LR}}$ is the celebrated Lieb-Robinson velocity. By substituting
Eqs.\,(\ref{eq:II-2}-\ref{eq:II-4}) into Eqs.\,(\ref{eq:Gam-sq}),
we find 
\begin{align}
\frac{1}{4}\Gamma^{2}(t) & \lesssim NR^{2}+NR^{3}+R^{2}\sum_{j=1}^{N}\sum_{k\notin\mathcal{B}(j,\,R)}\exp\left(-[d(X_{j},\,X_{k})-v_{\text{LR}}t]/\xi\right)\nonumber \\
 & \lesssim N+Ne^{v_{\text{LR}}t/\xi}\sum_{n=1}^{N}e^{-n/\xi}\nonumber \\
 & \lesssim N+Ne^{v_{\text{LR}}t/\xi}\int_{1}^{N}e^{-x/\xi}dx\nonumber \\
 & \lesssim N+Ne^{v_{{\rm LR}}t/\xi}\xi(e^{-1/\xi}-e^{-N/\xi})\nonumber \\
 & \leq\gamma^{2}(t)N,
\end{align}
where $\gamma(t)$ is a function of only time, independent of $N$,
and we have used the results $R\sim O(1)$ and $\lim_{N\to\infty}e^{-N/\xi}=0$.

\section{\label{sec:I-large-t}Proof of $\lim_{t\to+\infty}G(t)/t$ being
time-independent for time-independent Hamiltonians}

For time-independent Hamiltonians $H_{\lambda}(t)=H_{\lambda}$, the
generator for quantum sensing $\lambda$, Eq.\,(2) is given
by 
\begin{align}
G(t) & =\int_{0}^{t}e^{{\rm i}H_{\lambda}\tau}\partial_{\lambda}H_{\lambda}e^{-{\rm i}H_{\lambda}\tau}d\tau\nonumber \\
 & \overset{\tau=t(1-s)}{=}t\int_{0}^{1}e^{{\rm i}H_{\lambda}t(1-s)}\partial_{\lambda}H_{\lambda}e^{{\rm -i}H_{\lambda}t(1-s)}ds.
\end{align}
By using the Baker--Campbell--Hausdorff formula \citep{Hall2013},
the above equation can be rewritten as 
\begin{align}
\frac{G(t)}{t} & =\sum_{n=0}^{\infty}\int_{0}^{1}(1-s)^{n}ds\frac{({\rm i}t\mathcal{L})^{n}}{n!}\partial_{\lambda}H_{\lambda}\nonumber \\
 & =\sum_{n=0}^{\infty}\frac{({\rm i}t\mathcal{L})^{n}}{(n+1)!}\partial_{\lambda}H_{\lambda}\nonumber \\
 & \equiv g({\rm i}t\mathcal{L})\partial_{\lambda}H_{\lambda},\label{eq:G-over-t}
\end{align}
where the Liouvillian is defined as $\mathcal{L}(O)\equiv[H_{\lambda},O]$
whose eigenvalues are $L_{mn}=E_{m}-E_{n}$ and 
\begin{equation}
g(x)\equiv\sum_{n=0}^{\infty}\frac{x^{n}}{(n+1)!}=\frac{e^{x}-1}{x}.\label{eq:h-function}
\end{equation}
By the Choi-Jamio$\l$kowski isomorphism, we can rewrite operator
$|\partial_{\lambda}H_{\lambda})$ as a vector in the space $\mathcal{H\otimes\mathcal{H}}$
where $\mathcal{H}$ is the original Hilbert space. Due to the Hermiticity
of $\mathcal{L}$, we can assume that it can be diagonalized as $\mathcal{L}=\sum_{mn}L_{mn}|L_{mn})(L_{mn}|$
so that 
\begin{equation}
\frac{G(t)}{t}\equiv\sum_{mn}g({\rm i}tL_{mn})|L_{mn})(L_{mn}|\partial_{\lambda}H_{\lambda}).
\end{equation}
We observe that in the limit $t\to\infty$
\begin{equation}
\lim_{t\to\infty}g({\rm i}tL_{mn})=0,\,\text{for},\,L_{mn}\neq0.\label{eq:L-gap}
\end{equation}
Therefore, we conclude
\begin{equation}
\lim_{t\to+\infty}\frac{G(t)}{t}=\sum_{mn}\delta_{L_{mn},0}|L_{mn})(L_{mn}|\partial_{\lambda}H_{\lambda}).
\end{equation}
which is time-independent. Apparently, the condition to reach this
limit is 
\begin{equation}
t\gg1/\min_{m,\,n}\{|L_{mn}|>0\}.
\end{equation}
This regime is reached when time reaches the time scale of the inverse of the minimum non-zero energy difference of the system. Since $I(t)=\text{4Var}[G(t)]{}_{\ket{\psi_{0}}}$, we can assume $I(t)\sim t^{2}N^{\alpha}$ for sufficient long-time
and large $N$. 

\section{\label{sec:fast-decay-2-body}Proof of the SNL for fast-decaying
two-body metrological operator}

Then it is straightforward to verify that 
\begin{equation}
[\partial_{\lambda}H_{\lambda}(t)]^{(\text{H})}=\sum_{i}\sum_{j\ge i}\sum_{\alpha}\eta_{ij}^{\alpha}\mathcal{O}_{ij}^{\alpha}=\sum_{i}\sum_{j\leq i}\sum_{\alpha}\eta_{ji}^{\alpha}\mathcal{O}_{ji}^{\alpha}
\end{equation}
So one can find 
\begin{equation}
[\partial_{\lambda}H_{\lambda}(t)]^{(\text{H})}=\frac{1}{2}\sum_{i}\sum_{\alpha}\left(\sum_{j\ge i}\eta_{ij}^{\alpha}\mathcal{O}_{ij}^{\alpha}+\sum_{j\leq i}\sum_{\alpha}\eta_{ji}^{\alpha}\mathcal{O}_{ji}^{\alpha}\right)\equiv\sum_{i}\tilde{\mathscr{O}}_{i}
\end{equation}
where $\tilde{\mathscr{O}}_{i}$ is defined in the main text. To facilitate
the following discussions, we also introduce
\begin{equation}
\mathscr{O}_{i}\equiv\sum_{j\ge i}^{N}\sum_{\alpha}\eta_{ij}^{\alpha}\mathcal{O}_{ij}^{\alpha}.
\end{equation}
Then it is straightforward to show that 
\begin{equation}
\frac{1}{4}\Gamma^{2}(t)=\sum_{k}\text{Cov}[\mathscr{O}_{k}\mathscr{O}_{k}]{}_{\ket{\psi_{\text{sep}}}}+2\sum_{k}\sum_{i<k}\text{Cov}[\mathscr{O}_{i}\mathscr{O}_{k}]{}_{\ket{\psi_{\text{sep}}}},
\end{equation}
where $\ket{\psi_{{\rm sep}}}$ is a separable state. Thus, we find
\begin{equation}
\frac{1}{4}\Gamma^{2}(t)=\sum_{k}S_{N}(k),
\end{equation}
where $S_{N}(k)\equiv S_{N}^{(1)}(k)+S_{N}^{(2)}(k)$ and 
\begin{align}
S_{N}^{(1)}(k) & \equiv\text{\ensuremath{\sum_{\alpha\beta}}\ensuremath{\sum_{\ell\geq k}\sum_{j\geq k}\eta_{k\ell}^{\alpha}}\ensuremath{\eta_{kj}^{\beta}}Cov}[\mathcal{O}_{k\ell}^{\alpha}\mathcal{O}_{kj}^{\beta}]{}_{\ket{\psi_{\text{sep}}}},\nonumber \\
S_{N}^{(2)}(k) & =2\sum_{\alpha\beta}\sum_{i<k}\sum_{j\geq i}\sum_{\ell\geq k}\eta_{ij}^{\alpha}\eta_{k\ell}^{\beta}\text{Cov}[\mathcal{O}_{ij}^{\alpha}\mathcal{O}_{k\ell}^{\beta}]{}_{\ket{\psi_{\text{sep}}}}.\label{eq:SNK}
\end{align}
Our aim is to show that 
\begin{equation}
\lim_{N\to\infty}\frac{\Gamma^{2}(t)}{N}<\infty,\label{eq:Claim}
\end{equation}
which suggests that SNL cannot be surpassed by separable states. To
show Eq.\,(\ref{eq:Claim}), it suffices to demonstrate that 
\begin{equation}
\lim_{k\to\infty}\lim_{N\to\infty}\left[S_{N}^{(1)}(k)+S_{N}^{(2)}(k)\right]<\infty\text{,}\label{eq:Claim-2}
\end{equation}
Due to $\|\mathcal{O}_{jk}^{\alpha}\|\sim O(1),$ we have 
\begin{equation}
{\rm Var}[\mathcal{O}_{k\ell}^{\alpha}])_{\ket{\psi_{{\rm 0}}}}\sim\|\mathcal{O}_{jk}^{\alpha}\|^{2}\sim O(1),\label{eq:condition-2}
\end{equation}
 as $N\to\infty$ where $\ket{\psi_{0}}$ can be any state and not
restricted to separable states. Also, by using the Cauchy-Schwartz
inequality, we can show that as $N\to\infty$
\begin{equation}
\big|\text{Cov}[\mathcal{O}_{ij}^{\alpha}\mathcal{O}_{k\ell}^{\beta}]{}_{\ket{\psi_{\text{0}}}}\big|\leq\sqrt{\text{Var}[\mathcal{O}_{ij}^{\alpha}]_{\ket{\psi_{\text{0}}}}}\sqrt{\text{Var}[\mathcal{O}_{k\ell}^{\beta}]_{\ket{\psi_{\text{0}}}}}\sim O(1).\label{eq:condition11}
\end{equation}
Furthermore, we have the following condition to characterize the locality
of the fast decay operator:
\begin{equation}
\lim_{k\to\infty}\lim_{N\to\infty}C_{N}^{\alpha}(k)<\infty,\label{eq:condition-3}
\end{equation}
 where 
\begin{equation}
C_{N}^{\alpha}(k)\equiv\sum_{i\leq k}\sum_{j\ge k}|\eta_{ij}^{\alpha}|\label{eq:C-def}
\end{equation}
Apparently
\begin{equation}
\sum_{j\ge k}|\eta_{kj}^{\alpha}|\leq C_{N}^{\alpha}(k)\label{eq:single-sum-ineq}
\end{equation}
In the limit, $N\to\infty$ and $k\to\infty$, Euler-Maclaurin formula
allows us to rewrite the series on r.h.s. of Eq.~(\ref{eq:C-def})
in terms of a double integral 
\[
\lim_{k\to\infty}\lim_{N\to\infty}C_{N}^{\alpha}(k)=\text{\ensuremath{\lim_{k\to\infty}}}\int_{1}^{k}dx\int_{k}^{\infty}dy|\eta_{xy}^{\alpha}|.
\]
which is easier to evaluate than the double series. 

Now we are in a position to show that
\begin{align}
\lim_{k\to\infty}\lim_{N\to\infty}|S_{N}^{(1)}(k)| & \lesssim\ensuremath{\sum_{\alpha\beta}}\lim_{k\to\infty}\lim_{N\to\infty}\left(\ensuremath{\sum_{\ell\geq k}\sum_{j\geq k}|\eta_{k\ell}^{\alpha}}\ensuremath{\eta_{kj}^{\beta}}|\right)\nonumber \\
 & \lesssim\ensuremath{\sum_{\alpha\beta}}\lim_{k\to\infty}\lim_{N\to\infty}C_{N}^{\alpha}(k)C_{N}^{\beta}(k)\nonumber \\
 & <\infty,
\end{align}
where we have utilized Eq.\,(\ref{eq:condition11}) in the first
inequality, Eq.~(\ref{eq:single-sum-ineq}) in the second inequality,
and the fact that the number of terms of the sum over $\alpha$ is
finite in the last inequality. Next, we will discuss two cases where
the $\mathcal{O}$ is two-body spin or fermionic operators, respectively,
to show that 
\begin{equation}
\lim_{k\to\infty}\lim_{N\to\infty}|S_{N}^{(2)}(k)|<\infty.\label{eq:S2-abs}
\end{equation}

\subsection{Two-body spin operators}

When $\mathcal{O}_{ji}^{\alpha}$ and $\mathcal{O}_{kl}^{\beta}$
are two-body spin operators, we note that if $\{j,\,i\}$ does not
overlap with $\{k,\,l\}$ then the correlation is zero since $\ket{\psi_{{\rm sep}}}$
is a separable state. With this observation, we find
\begin{align}
|S_{N}^{(2)}(k)| & =2\sum_{\alpha\beta}\sum_{i<k}\sum_{j\geq i}\sum_{\ell\geq k}|\eta_{ij}^{\alpha}\eta_{k\ell}^{\beta}\text{Cov}[\mathcal{O}_{ij}^{\alpha}\mathcal{O}_{k\ell}^{\beta}]{}_{\ket{\psi_{\text{sep}}}}\delta_{k,j}|\nonumber \\
 & =2\sum_{\alpha\beta}\sum_{i<k}\sum_{\ell\geq k}|\eta_{ik}^{\alpha}\eta_{k\ell}^{\beta}\text{Cov}[\mathcal{O}_{ik}^{\alpha}\mathcal{O}_{k\ell}^{\beta}]{}_{\ket{\psi_{\text{sep}}}}|\nonumber \\
 & \lesssim\ensuremath{\sum_{\alpha\beta}}\sum_{i<k}\sum_{\ell\geq k}|\eta_{ik}^{\alpha}\eta_{k\ell}^{\beta}|,
\end{align}
where the appearance of $\delta_{k,j}$ in the first equality comes
from the initial state is restricted to separable states. Using (\ref{eq:single-sum-ineq})
,

\begin{align*}
|S_{N}^{(2)}(k)| & \lesssim\ensuremath{\sum_{\alpha\beta}}C_{N}^{\alpha}(k)C_{N}^{\beta}(k),
\end{align*}
Taking the limit $N\to\infty$ and $k\to\infty$, we conclude Eq.~(\ref{eq:S2-abs}). 

If we do not require separable initial states then 
\begin{align}
|S_{N}^{(2)}(k)| & =2\sum_{\alpha\beta}\sum_{i<k}\sum_{j\geq i}\sum_{\ell\geq k}|\eta_{ij}^{\alpha}\eta_{k\ell}^{\beta}\text{Cov}[\mathcal{O}_{ij}^{\alpha}\mathcal{O}_{k\ell}^{\beta}]{}_{\ket{\psi_{\text{0}}}}|\nonumber \\
 & \lesssim\ensuremath{\sum_{\alpha\beta}}\sum_{i<k}\sum_{j\geq i}\sum_{\ell\geq k}|\eta_{ij}^{\alpha}\eta_{k\ell}^{\beta}|\nonumber \\
 & \lesssim\sum_{\alpha\beta}C^{\beta}_N(k)\sum_{i<k}\sum_{j\geq i}|\eta_{ij}^{\alpha}|,
\end{align}
which suggests that $\lim_{k\to\infty}\lim_{N\to\infty}S_{N}^{(2)}(k)<\infty$
is not guaranteed since $\sum_{i<k}\sum_{j\geq i}|\eta_{ij}^{\alpha}|$
may not be bounded given Eqs.\,(\ref{eq:condition-3}) and (\ref{eq:C-def}).
Actually, the HL can be achieved for GHZ-like entangled initial states
and thus Eq.\,(\ref{eq:Claim-2}) is not expected to hold.

\subsection{Two-body fermionic operators}

When $\mathcal{O}_{ij}^{\alpha}$ and $\mathcal{O}_{kl}^{\beta}$
are two-body fermionic operators, see the below: 

\begin{align}
c_{i}^{\dagger}c_{j} & =\left\{ \begin{array}{cc}
\sigma_{i}^{-}\sigma_{i+1}^{z}\sigma_{i+2}^{z}\cdots\sigma_{j-1}^{z}\sigma_{j}^{+}, & i<j,\\
(1-\sigma_{i}^{z})/2, & i=j,\\
\sigma_{j}^{+}\sigma_{j+1}^{z}\sigma_{j+2}^{z}\cdots\sigma_{i-1}^{z}\sigma_{i}^{-}, & i>j.
\end{array}\right.\nonumber \\
c_{i}c_{j}^{\dagger} & =\left\{ \begin{array}{cc}
-\sigma_{i}^{+}\sigma_{i+1}^{z}\sigma_{i+2}^{z}\cdots\sigma_{j-1}^{z}\sigma_{j}^{-}, & i<j,\\
(1+\sigma_{i}^{z})/2, & i=j,\\
-\sigma_{j}^{-}\sigma_{j+1}^{z}\sigma_{j+2}^{z}\cdots\sigma_{i-1}^{z}\sigma_{i}^{+}, & i>j.
\end{array}\right.\\
c_{i}^{\dagger}c_{j}^{\dagger} & =\left\{ \begin{array}{cc}
\sigma_{i}^{-}\sigma_{i+1}^{z}\sigma_{i+2}^{z}\cdots\sigma_{j-1}^{z}\sigma_{j}^{-}, & i<j,\\
0, & i=j,\\
-\sigma_{j}^{-}\sigma_{j+1}^{z}\sigma_{j+2}^{z}\cdots\sigma_{i-1}^{z}\sigma_{i}^{-}, & i>j,
\end{array}\right.\label{eq:Num-4}
\end{align}
which suggests that if $[i,j]\cap[k,\ell]=\text{\ensuremath{\varnothing}}$
then ${\rm Cov}[\mathcal{O}_{ji}^{\alpha}\mathcal{O}_{k\ell}^{\beta}]_{\ket{\psi_{{\rm sep}}}}=0.$
Thus, we can find

\begin{align}
S_{N}^{(k)}(k) & =2\sum_{\alpha\beta}\sum_{i<k}\sum_{j\geq i}\sum_{\ell\geq k}\eta_{ij}^{\alpha}\eta_{k\ell}^{\beta}\text{Cov}[\mathcal{O}_{ij}^{\alpha}\mathcal{O}_{k\ell}^{\beta}]{}_{\ket{\psi_{\text{sep}}}}
\end{align}
Using the condition that the initial state is separable, one can replace
$j\ge i$ in above summation with $j\ge k$, leading to 
\begin{align}
S_{N}^{(2)}(k) & =2\sum_{\alpha\beta}\sum_{i<k}\sum_{j\geq k}\sum_{\ell\geq k}\eta_{ij}^{\alpha}\eta_{k\ell}^{\beta}\text{Cov}[\mathcal{O}_{ij}^{\alpha}\mathcal{O}_{k\ell}^{\beta}]{}_{\ket{\psi_{\text{sep}}}}
\end{align}
Thus, we find 
\[
|S_{N}^{(2)}(k)|\lesssim\sum_{\alpha\beta}\sum_{i<k}\sum_{j\geq k}\sum_{\ell\geq k}|\eta_{ij}^{\alpha}\eta_{k\ell}^{\beta}|\leq\sum_{\alpha\beta}C_{N}^{\alpha}(k)\sum_{\ell\geq k}|\eta_{k\ell}^{\beta}|\lesssim\ensuremath{\sum_{\alpha\beta}}C_{N}^{\alpha}(k)C_{N}^{\beta}(k).
\]
Taking $N\to\infty$ and $k\to\infty$ and using Eq.\,(\ref{eq:condition-3}),
we prove Eq.~(\ref{eq:S2-abs}).

\section{\label{sec:G-TFI}Metrological generator for the TFI periodic chain}

The integrable Ising model Eq.~(14)
can be diagonalized as a free fermion model \citep{LIEB1961407,PFEUTY197079}.
Defining $\sigma_{i}^{\pm}=(\sigma_{i}^{x}\pm{\rm i}\sigma_{i}^{y})/2$,
we can construct fermion creation and annihilation operators by the
Jordan-Wigner transformation: 
\begin{equation}
c_{i}^{\dagger}=\prod_{j=1}^{i-1}\sigma_{j}^{z}\sigma_{i}^{-}.\label{eq:JW trans}
\end{equation}
The inverse transformation is then given by 
\begin{equation}
\sigma_{i}^{z}=1-2c_{i}^{\dagger}c_{i},\quad\sigma_{i}^{x}=\prod_{j=1}^{i-1}(1-2c_{j}^{\dagger}c_{j})(c_{i}+c_{i}^{\dagger}).\label{eq:inverse JW trans}
\end{equation}
In terms of fermionic operators, the Hamiltonian can be expressed
as 
\begin{align}
H_{\lambda}^{{\rm TFI}} & =-J\sum_{i=1}^{N}\sigma_{i}^{x}\cdot\sigma_{i+1}^{x}-\lambda\sum_{i=1}^{N}\sigma_{i}^{z}\nonumber \\
 & =-J\sum_{i=1}^{N-1}(c_{i}^{\dagger}-c_{i})(c_{i+1}^{\dagger}+c_{i+1})-\lambda\sum_{i=1}^{N}(1-2c_{i}^{\dagger}c_{i})+JP(c_{N}^{\dagger}-c_{N})(c_{1}^{\dagger}+c_{1}),
\end{align}
where $P=\exp({\rm i}\pi\sum_{i=1}^{N}c_{i}^{\dagger}c_{i})$ is
the parity operator commuting with the Hamiltonian. 
Thus, the above
Hamiltonian can be rewritten according to the parity symmetry, i.e.,
\[
H_{\lambda}^{{\rm TFI}}=\text{(}H_{\lambda}^{{\rm TFI}})_{{\rm odd}}\oplus\text{(}H_{\lambda}^{{\rm TFI}})_{{\rm even}},
\]
where ``odd'' and ``even'' correspond to the Hamiltonian acting
on the subspaces of the Fock space with an odd or even number of fermions,
which is also related with periodic ($P=-1$) or antiperiodic ($P=1$)
boundary conditions. 

It can be shown that $\text{(}H_{\lambda}^{{\rm TFI}})_{{\rm odd}}=(H_{\lambda}^{{\rm TFI}})_{{\rm even}}$
at the thermodynamic limit $N\to\infty$. Thus, for convenience, we
work on the subspace with an even number of fermions. 
Introducing
the Fourier transformation
\begin{equation}
f_{k}=\frac{1}{\sqrt{N}}\sum_{j=1}^{N}c_{j}{\rm e}^{{\rm i}kj},\quad k=\frac{2\pi(n+1/2)}{N},\quad n=-\frac{N}{2},-\frac{N}{2}+1,\ldots,\frac{N}{2}-1,\label{eq:Fourier trans}
\end{equation}
and the Bogoliubov fermions
\begin{equation}
f_{k}=\cos\text{\ensuremath{\frac{\theta_{k}}{2}}}\alpha_{k}+{\rm i}\sin\frac{\theta_{k}}{2}\alpha_{-k}^{\dagger},\label{eq:Bogoliubov Fermions}
\end{equation}
the Hamiltonian reduces to the diagonalized form 
\begin{equation}
H_{\lambda}^{{\rm TFI}}=\sum_{k}\epsilon_{k}\alpha_{k}^{\dagger}\alpha_{k},
\end{equation}
where $\epsilon_{k}=2\sqrt{J^{2}+\lambda^{2}-2J\lambda\cos k}$ and
$\theta_{k}$ is determined by
\begin{equation}
\sin\theta_{k}=\frac{-2J\sin k}{\epsilon_{k}},\quad\cos\theta_{k}=\frac{2(\lambda-J\cos k)}{\text{\ensuremath{\epsilon_{k}}}}.\label{eq:Num-9}
\end{equation}
In terms of $\alpha_{k}^{\dagger}$ and $\alpha_{k}$, $\partial_{\lambda}H_{\lambda}^{{\rm TFI}}=\sum_{i=1}^{N}\sigma_{i}^{z}$
can be expressed as 
\begin{equation}
\partial_{\lambda}H_{\lambda}^{{\rm TFI}}=N-2\sum_{k}\left(\cos^{2}\frac{\theta_{k}}{2}\alpha_{k}^{\dagger}\alpha_{k}+\sin^{2}\frac{\theta_{k}}{2}\alpha_{k}\alpha_{k}^{\dagger}+\frac{{\rm i}}{2}\sin\theta_{k}\alpha_{k}^{\dagger}\alpha_{-k}^{\dagger}-\frac{{\rm i}}{2}\sin\theta_{k}\alpha_{-k}\alpha_{k}\right).
\end{equation}
Thus, we obtain

\begin{align}
[\partial_{\lambda}H_{\lambda}^{{\rm TFI}}(\tau)]^{{\rm (H)}} & =U^{\dagger}(\tau)\partial_{\lambda}H_{\lambda}^{{\rm TFI}}U(\tau)\nonumber \\
 & =N-2\sum_{k}\left(\cos^{2}\frac{\theta_{k}}{2}\alpha_{k}^{\dagger}\alpha_{k}+\sin^{2}\frac{\theta_{k}}{2}\alpha_{k}\alpha_{k}^{\dagger}+\frac{{\rm i}}{2}{\rm e}^{2{\rm i}\epsilon_{k}\tau}\sin\theta_{k}\alpha_{k}^{\dagger}\alpha_{-k}^{\dagger}-\frac{{\rm i}}{2}{\rm e}^{-2{\rm i}\epsilon_{k}\tau}\sin\theta_{k}\alpha_{-k}\alpha_{k}\right).\label{eq:Generator-motion}
\end{align}
By using Eq.\,(\ref{eq:Bogoliubov Fermions}), the Eq.\,(\ref{eq:Generator-motion})
can be expressed as 
\begin{equation}
[\partial_{\lambda}H_{\lambda}^{{\rm TFI}}(\tau)]^{{\rm (H)}}=N-\sum_{k}\left[A(k)f_{k}^{\dagger}f_{k}+B(k)f_{k}f_{k}^{\dagger}+D(k)f_{k}^{\dagger}f_{-k}^{\dagger}+D^{*}(k)f_{-k}f_{k}\right],\label{eq:Generator-motion-fk}
\end{equation}
where
\begin{align}
A(k) & =1+\cos^{2}\theta_{k}+\sin^{2}\theta_{k}\cos(2\epsilon_{k}\tau),\nonumber \\
B(k) & =1-\cos^{2}\theta_{k}-\sin^{2}\theta_{k}\cos(2\epsilon_{k}\tau),\nonumber \\
D(k) & ={\rm i}\sin\theta_{k}\cos\theta_{k}[\cos(2\epsilon_{k}\tau)-1]-\sin\theta_{k}\sin(2\epsilon_{k}\tau).
\end{align}
By further using the Fourier transformation (\ref{eq:Fourier trans}),
we obtain 
\begin{equation}
[\partial_{\lambda}H_{\lambda}^{{\rm TFI}}]^{{\rm (H)}}=\sum_{i=1}^{N}\mathscr{O}_{i}(\tau),
\end{equation}
where
\begin{equation}
\mathscr{O}_{i}(\tau)=1-\sum_{j=1}^{N}\left[\tilde{A}(j-i)c_{i}^{\dagger}c_{j}+\tilde{B}(i-j)c_{i}c_{j}^{\dagger}+\tilde{D}(j-i)c_{i}^{\dagger}c_{j}^{\dagger}+\tilde{D^{*}}(j-i)c_{i}c_{j}\right],\label{eq:O_ising}
\end{equation}
and $\tilde{A}(\ell)$ is the Fourier transformation of the function
$A(k)$, i.e., 
\begin{equation}
\tilde{A}(\ell)\equiv\frac{1}{N}\sum_{k}A(k){\rm e}^{{\rm i}k\ell}.\label{eq:tilde_A}
\end{equation}

We can further rewrite Eq.\,(\ref{eq:O_ising}) in terms of Eq.\,(12)
in the main text, i.e.,
\begin{equation}
\mathscr{O}_{i}=\sum_{j\ge i}\sum_{\alpha}\eta_{ij}^{\alpha}\mathcal{O}_{ij}^{(\alpha)}
\end{equation}
where we have suppressed the time-dependence, $\mathcal{O}_{ij}^{(1)}=(c_{i}^{\dagger}c_{j}+{\rm h.c.})$,
$\mathcal{O}_{ij}^{(2)}=(c_{i}c_{j}^{\dagger}+{\rm h.c.)}$, $\mathcal{O}_{ij}^{(3)}=(c_{i}^{\dagger}c_{j}^{\dagger}+{\rm h.c.}),$
and $O_{ij}^{(4)}={\rm i}(c_{i}^{\dagger}c_{j}^{\dagger}-{\rm h.c.})$.
The corresponding $\eta$-functions are given by 
\begin{align}
\eta_{ij}^{(1)} & =-(1-\delta_{ij})\tilde{A}(j\!-\!i)-\delta_{ij}\frac{\tilde{A}(0)}{2},\nonumber \\
\eta_{ij}^{(2)} & =-(1-\delta_{ij})\tilde{B}(j\!-\!i)-\delta_{ij}\frac{\tilde{B}(0)}{2},\nonumber \\
\eta_{ij}^{(3)} & =-2{\rm Re}[\tilde{D}(j\text{\!}-\!i)],\nonumber \\
\eta_{ij}^{(4)} & =-2{\rm Im}[\tilde{D}(j\!-\!i)].
\end{align}
The expressions for $\tilde{A}(\ell)$, $\tilde{B}(\ell)$, $\tilde{C}(\ell)$
and $\tilde{D}(\ell)$ are dramatically simplified in the limit $N\to\infty$
and $t\to\infty$. In the limit, $N\to\infty$, we can rewrite $\tilde{A}(\ell)$
as an integral
\begin{align}
\tilde{A}(\ell) & =\frac{1}{2\pi}\int_{-\pi}^{\pi}A(k)e^{ik\ell}dk\nonumber \\
 & =\delta_{\ell,0}+\frac{1}{2\pi}\int_{-\pi}^{\pi}\cos^{2}\theta_{k}e^{ik\ell}dk+\frac{1}{2\pi}\int_{-\pi}^{\pi}\sin^{2}\theta_{k}\cos(2\epsilon_{k}\tau)e^{ik\ell}dk,\label{eq:a1}
\end{align}
By the Riemann-Lebesgue lemma~\citep{bender1999advanced}, the last
integral in the right hand of Eq.\,(\ref{eq:a1}) tends to zero as
$\tau\to\infty$. The second integral in the right hand of Eq.\,(\ref{eq:a1})
can be evaluated by introducing $z=e^{ik}$:
\begin{equation}
I(\ell)\equiv\frac{1}{2\pi}\int_{-\pi}^{\pi}\cos^{2}\theta_{k}e^{ik\ell}dk=\frac{1}{2\pi{\rm i}}\int_{\mathscr{C}}\frac{\left(J\frac{z^{2}+1}{2}-\lambda z\right)^{2}}{(J^{2}+\lambda^{2})z-J\lambda(z^{2}+1)}z^{\ell-2}dz.
\end{equation}
where the contour $\mathscr{C}$ is along the counterclockwise direction
of the unit circle on the complex plane.

We denote $I(\ell)=\frac{1}{2\pi{\rm i}}\int_{\mathscr{C}}f(z)dz$,
where 
\begin{equation}
f(z)\equiv\frac{\left(J\frac{z^{2}+1}{2}-\lambda z\right)^{2}}{(J^{2}+\lambda^{2})z-J\lambda(z^{2}+1)}z^{\ell-2}
\end{equation}
The residues of $f$ are listed as follows:
\begin{align}
{\rm Res}(f,0) & =\frac{2\lambda^{2}-J^{2}}{4\lambda^{2}}\quad{\rm for}\quad\ell=0;\nonumber \\
{\rm Res}(f,0) & =-\frac{J}{4\lambda}\quad{\rm for\quad\ell=1;}\nonumber \\
{\rm Res}\left(f,\frac{J}{\lambda}\right) & =\left(\frac{J}{\lambda}\right)^{\ell}\frac{\lambda^{2}-J^{2}}{4\lambda^{2}},\quad\text{for}\quad\text{all}\quad\ell\nonumber \\
{\rm Res}\left(f,\frac{\lambda}{J}\right) & =\left(\frac{\lambda}{J}\right)^{\ell-2}\frac{J^{2}-\lambda^{2}}{4J^{2}},\quad\text{for}\quad\text{all}\quad\ell.
\end{align}
Then, by using the residue theorem, we obtain $I(\ell)={\rm Res}(f,0)$
for $J=\lambda$, $I(\ell)={\rm Res}(f,0)+{\rm Res}\left(f,\frac{\lambda}{J}\right)$
for $J>\lambda>0$, and $I(\ell)={\rm Res}(f,0){\rm +Res}\left(f,\frac{J}{\lambda}\right)$
for $0<J<\lambda$. Finally, we have
\begin{equation}
\tilde{A}(\ell)=\left\{ \begin{array}{cc}
3/2, & \ell=0,\\
-1/4 & \ell=1,\\
0, & \ell=2,3,\ldots,N-1,
\end{array}\right.\quad{\rm for}\quad J=\lambda>0;
\end{equation}

\begin{equation}
\tilde{A}(\ell)=\left\{ \begin{array}{cc}
3/2, & \ell=0,\\
-\lambda/4J & \ell=1,\\
\frac{J^{2}-\lambda^{2}}{4J^{2}}\left(\frac{\text{\ensuremath{\lambda}}}{J}\right)^{\ell-2}, & \ell=2,3,\ldots,N-1,
\end{array}\right.\quad{\rm for}\quad J>\lambda>0;
\end{equation}

\begin{equation}
\tilde{A}(\ell)=\left\{ \begin{array}{cc}
2-\frac{J^{2}}{2\lambda^{2}}, & \ell=0,\\
-\frac{J^{3}}{4\lambda^{3}} & \ell=1,\\
\frac{\lambda^{2}-J^{2}}{4\lambda^{2}}\left(\frac{\text{\ensuremath{J}}}{\lambda}\right)^{\ell}, & \ell=2,3,\ldots,N-1,
\end{array}\right.\quad{\rm for}\quad0<J<\lambda.
\end{equation}
Similarly, we can obtain the other terms under the limit $\tau\to\infty$
and $N\to\infty$
\begin{align}
\tilde{B}(\ell) & =\left\{ \begin{array}{cc}
1/2, & \ell=0,\\
1/4, & \ell=1,\\
0, & \ell=2,3,\ldots,N-1,
\end{array}\right.\quad{\rm for}\quad J=\lambda>0,\quad\tilde{B}(\ell)=\left\{ \begin{array}{cc}
1/2, & \ell=0,\\
\lambda/4J, & \ell=1,\\
-\frac{J^{2}-\lambda^{2}}{4J^{2}}\left(\frac{\text{\ensuremath{\lambda}}}{J}\right)^{\ell-2}, & \ell=2,3,\ldots,N-1,
\end{array}\right.\quad{\rm for}\quad J>\lambda>0,\nonumber \\
\tilde{B}(\ell) & =\left\{ \begin{array}{cc}
\frac{J^{2}}{2\lambda^{2}}, & \ell=0,\\
\frac{J^{3}}{4\lambda^{3}}, & \ell=1,\\
-\frac{\lambda^{2}-J^{2}}{4\lambda^{2}}\left(\frac{\text{\ensuremath{J}}}{\lambda}\right)^{\ell}, & \ell=2,3,\ldots,N-1,
\end{array}\right.\quad{\rm for}\quad0<J<\lambda,
\end{align}
\begin{align}
\tilde{D}(\ell) & =\left\{ \begin{array}{cc}
0, & \ell=0,\\
-1/4 & \ell=1,\\
0, & \ell=2,3,\ldots,N-1,
\end{array}\right.\quad{\rm for}\quad J=\lambda>0,\quad\tilde{D}(\ell)=\left\{ \begin{array}{cc}
0, & \ell=0,\\
-\lambda/4J & \ell=1,\\
\frac{J^{2}-\lambda^{2}}{4J^{2}}\left(\frac{\text{\ensuremath{\lambda}}}{J}\right)^{\ell-2}, & \ell=2,3,\ldots,N-1,
\end{array}\right.\quad{\rm for}\quad J>\lambda>0,\nonumber \\
\tilde{D}(\ell) & =\left\{ \begin{array}{cc}
0, & \ell=0,\\
\frac{J^{3}-2J\lambda^{2}}{4\lambda^{3}} & \ell=1,\\
\frac{J^{2}-\lambda^{2}}{4\lambda^{2}}\left(\frac{\text{\ensuremath{J}}}{\lambda}\right)^{\ell}, & \ell=2,3,\ldots,N-1,
\end{array}\right.\quad{\rm for}\quad0<J<\lambda,
\end{align}
and $\tilde{D^{*}}(\ell)=-\tilde{D}(\ell)$. 

\section{\label{sec:QFI-TFI}QFI for the TFI model}

According to Eq.\,(\ref{eq:Generator-motion}), the generator $G(t)$
given in Eq.\,(2) of the integrable Ising model is given
by 
\begin{align}
\frac{G(t)}{t} & =\frac{1}{t}\int_{0}^{t}\left[\partial_{\lambda}H_{\lambda}^{{\rm TFI}}(\tau)\right]^{(\text{H})}d\tau\nonumber \\
 & =N-2\sum_{k}\left(\cos^{2}\frac{\theta_{k}}{2}\alpha_{k}^{\dagger}\alpha_{k}+\sin^{2}\frac{\theta_{k}}{2}\alpha_{k}\alpha_{k}^{\dagger}+\frac{{\rm i}}{2}g(2{\rm i}\epsilon_{k}t)\sin\theta_{k}\alpha_{k}^{\dagger}\alpha_{-k}^{\dagger}-\frac{{\rm i}}{2}g(-2{\rm i}\epsilon_{k}t)\sin\theta_{k}\alpha_{-k}\alpha_{k}\right),\label{eq:Num-1}
\end{align}
where function $g(x)$ is given in Eq.\,(\ref{eq:h-function}). 

For the numerical calculation of QFI, we need to express $G(t)$ in
the spin representation. Thus, by using Eq.\,(\ref{eq:Bogoliubov Fermions})
and (\ref{eq:Fourier trans}), we can rewrite Eq.\,(\ref{eq:Num-1})
as
\begin{align}
\text{\ensuremath{\frac{G(t)}{t}}} & =N-\sum_{k}\left[\mathcal{A}(k)f_{k}^{\dagger}f_{k}+\mathcal{B}(k)f_{k}f_{k}^{\dagger}+\mathcal{D}(k)f_{k}^{\dagger}f_{-k}^{\dagger}+\mathcal{D}^{*}(k)f_{-k}f_{k}\right]\nonumber \\
 & =N-\sum_{i,j=1}^{N}\left[\tilde{\mathcal{A}}(j-i)c_{i}^{\dagger}c_{j}+\tilde{\mathcal{B}}(i-j)c_{i}c_{j}^{\dagger}+\tilde{\mathcal{D}}(j-i)c_{i}^{\dagger}c_{j}^{\dagger}+\tilde{\mathcal{D}^{*}}(j-i)c_{i}c_{j}\right]\label{eq:Num-2}
\end{align}
where 
\begin{align}
\mathcal{A}(k) & =1+\cos^{2}\theta_{k}+\frac{1}{2}\sin^{2}\theta_{k}[g(2{\rm i}\epsilon_{k}t)+g(-2{\rm i}\epsilon_{k}t)],\nonumber \\
\mathcal{B}(k) & =1-\cos^{2}\theta_{k}-\frac{1}{2}\sin^{2}\theta_{k}[g(2{\rm i}\epsilon_{k}t)+g(-2{\rm i}\epsilon_{k}t)],\nonumber \\
\mathcal{D}(k) & ={\rm i}\sin\theta_{k}\cos\theta_{k}\left[\frac{1}{2}g(2{\rm i}\epsilon_{k}t)+\frac{1}{2}g(-2{\rm i}\epsilon_{k}t)-1\right]+\frac{{\rm i}}{2}\sin\theta_{k}[g(2{\rm i}\epsilon_{k}t)-g(-2{\rm i}\epsilon_{k}t)],\label{eq:Num-3}
\end{align}
and $\tilde{\mathcal{A}}(\ell)$ is the Fourier transformation of
the function $\mathcal{A}(k)$ like Eq.\,(\ref{eq:tilde_A}). Equations
(\ref{eq:Num-2}), (\ref{eq:Num-3}), and (\ref{eq:Num-4}) provide
a numerical approach to express $G(t)$ in terms of the spin operators.
Finally, QFI can be easily calculated by using $I(t)=\text{4Var}[G(t)]{}_{\ket{\psi_{0}}}$,
Eq.\,(3).

Now we consider a special case where the initial state $\ket{\psi_{0}}$
is chosen to be the paramagnetic product state $\ket{\uparrow\uparrow\cdots\uparrow}$,
which can be viewed as the ground state of the Ising chain $H_{\lambda_{*}}^{{\rm TFI}}$
(14) by taking $\lambda_{*}\to+\infty$
and we denote it as $\ket{g(\lambda_{*})}$. We denote the Bogoliubov
fermionic operators and the Bogoliubov angle of $H_{\lambda_{*}}^{{\rm TFI}}$
by $\alpha_{*k}$ and $\theta_{*k}$, respectively. We now consider
the QFI for the evolved state $U(t)\ket{g(\lambda_{*})}$ with respect
to the parameter $\lambda$ where $U(t)=\exp(-{\rm i}H_{\lambda}^{{\rm TFI}}t)$.
Here we focus on the long-time limit and thus Eq.\,(\ref{eq:Num-1})
reduces to 
\begin{align}
\lim_{t\to+\infty}\frac{G(t)}{t} & =N-2\sum_{k}\left(\cos^{2}\frac{\theta_{k}}{2}\alpha_{k}^{\dagger}\alpha_{k}+\sin^{2}\frac{\theta_{k}}{2}\alpha_{k}\alpha_{k}^{\dagger}\right)\nonumber \\
 & =-2\sum_{k}\cos\theta_{k}n_{k},\label{eq:Num-5}
\end{align}
where $n_{k}\equiv\alpha_{k}^{\dagger}\alpha_{k}$ and we use the
result $\lim_{t\to+\infty}g(\pm2{\rm i}\epsilon_{k}t)=0$ for $J\neq\lambda$.
A constant is ignored in the last equation since it does not affect
the value of QFI. It can be checked that the ground state $\ket{g(\text{\ensuremath{\lambda_{*}}})}$
can be expressed as 
\begin{equation}
\ket{g(\lambda_{*})}=\frac{1}{\mathcal{N}}\exp\left[{\rm i}\sum_{k>0}R(k)\alpha_{-k}^{\dagger}\alpha_{k}^{\dagger}\right]\ket{g(\lambda)},\label{eq:Num-6}
\end{equation}
where $\mathcal{N}$ is a normalization factor, $\ket{g(\lambda)}$
is the ground state of Hamiltonian $H_{\lambda}^{{\rm TFI}}$, and
\begin{equation}
R(k)=\tan\left(\frac{\theta_{k}-\theta_{*k}}{2}\right).\label{eq:Num-7}
\end{equation}
Then we have 
\begin{align}
\lim_{t\to+\infty}\frac{\langle G^{2}(t)\rangle}{t^{2}} & =4\sum_{k}\cos^{2}\theta_{k}\langle n_{k}^{2}+n_{k}n_{-k}\rangle+4\sum_{k}\sum_{k'(\neq\pm k)}\cos\theta_{k}\cos\theta_{k'}\langle n_{k}n_{k'}\rangle\nonumber \\
 & =4\sum_{k}\cos^{2}\theta_{k}\langle n_{k}^{2}+n_{k}n_{-k}\rangle+4\sum_{k}\sum_{k'(\neq\pm k)}\cos\theta_{k}\cos\theta_{k'}\langle n_{k}\rangle\langle n_{k'}\rangle\nonumber \\
 & =8\sum_{k}\cos^{2}\theta_{k}\frac{R^{2}(k)}{1+R^{2}(k)}+4\sum_{k}\sum_{k'(\neq\pm k)}\cos\theta_{k}\cos\theta_{k'}\frac{R^{2}(k)R^{2}(k')}{[1+R^{2}(k)][1+R^{2}(k')]},\nonumber \\
\lim_{t\to+\infty}\frac{\langle G(t)\rangle^{2}}{t^{2}} & =4\sum_{k,k'}\cos\theta_{k}\cos\theta_{k'}\langle n_{k}\rangle\langle n_{k'}\rangle\nonumber \\
 & =4\sum_{k,k'}\cos\theta_{k}\cos\theta_{k'}\frac{R^{2}(k)R^{2}(k')}{[1+R^{2}(k)][1+R^{2}(k')]},\label{eq:Num-8}
\end{align}
where $\langle O\rangle\equiv\langle g(\lambda_{*})|O|g(\lambda_{*})\rangle$
we use the results $\langle n_{k}\rangle=\langle n_{k}n_{-k}\rangle=R^{2}(k)/[1+R^{2}(k)]$.
Substituting Eqs.\,(\ref{eq:Num-9}), (\ref{eq:Num-7}), and (\ref{eq:Num-8})
into the definition of QFI, we obtain 
\begin{align}
\lim_{t\to+\infty}\frac{I(t)}{t^{2}} & =\lim_{t\to+\infty}4\frac{\langle G^{2}(t)\rangle-\langle G(t)\rangle^{2}}{t^{2}}\nonumber \\
 & =32\sum_{k}\cos^{2}\theta_{k}\frac{R^{2}(k)}{1+R^{2}(k)}\left[1-\frac{R^{2}(k)}{1+R^{2}(k)}\right]\nonumber \\
 & =8\sum_{k}\cos^{2}\theta_{k}(\sin\theta_{k}\cos\theta_{*k}-\cos\theta_{k}\sin\theta_{*k})^{2}\nonumber \\
 & =8J^{2}(\lambda-\lambda_{*})^{2}\sum_{k}\frac{(\lambda-J\cos k)^{2}\sin^{2}k}{(J^{2}+\lambda^{2}-2\lambda J\cos k)^{2}(J^{2}+\lambda_{*}{}^{2}-2\lambda_{*}J\cos k)}.\label{eq:Num-10}
\end{align}
In the thermodynamic limit $N\to\infty$, the above equation becomes
\begin{align}
\lim_{t,N\to+\infty}\frac{I(t)}{Nt^{2}} & =8J^{2}(\lambda-\lambda_{*})^{2}\times\frac{1}{2\pi}\int_{-\pi}^{\pi}\frac{(\lambda-J\cos k)^{2}\sin^{2}k}{(J^{2}+\lambda^{2}-2\lambda J\cos k)^{2}(J^{2}+\lambda_{*}^{2}-2\lambda_{*}J\cos k)}dk\nonumber \\
 & =8J^{2}(\lambda-\lambda_{*})^{2}\times\frac{1}{2\pi}\int_{\mathscr{C}}\frac{\left(\lambda-J\frac{z^{2}+1}{2z}\right)^{2}\left[1-\left(\frac{z^{2}+1}{2z}\right)^{2}\right]}{\left(J^{2}+\lambda^{2}-2\lambda J\frac{z^{2}+1}{2z}\right)^{2}\left(J^{2}+\lambda_{*}^{2}-2\lambda_{*}J\frac{z^{2}+1}{2z}\right)}\frac{dz}{iz},\label{eq:Num-11}
\end{align}
where the contour $\mathscr{C}$ is the the unit circle on the complex
plane and $z=e^{ik}$. Finally, by using the residue theorem, we obtain
\begin{equation}
\lim_{t,N\to\infty}\frac{I(t)}{Nt^{2}}=\left\{ \begin{array}{cc}
\frac{(\lambda-\lambda{}_{*})}{(J^{2}-\lambda\lambda_{*})^{2}}(J^{2}-2\lambda\lambda_{*}+\lambda{}_{*}^{2}), & 0<\lambda,\lambda_{*}<J,\\
\frac{J^{2}(\lambda-\lambda_{*})^{2}}{\lambda_{*}{}^{2}\lambda^{3}(J^{2}-\lambda\lambda_{*})^{2}}(\lambda J^{4}+2\lambda_{*}J^{4}-4\lambda^{2}\lambda_{*}J^{2}-3\lambda\lambda_{*}J^{2}+4\lambda^{3}\lambda_{*}{}^{2}), & J<\lambda,\lambda_{*},\\
\frac{1}{\lambda^{3}}[2\lambda_{*}J^{2}-3\lambda J^{2}+\lambda(\lambda_{*}-2\lambda)^{2}], & 0<\lambda_{*}<J<\lambda,\\
\frac{1}{\lambda_{*}{}^{2}}(J^{2}-2\lambda\lambda_{*}+\lambda_{*}{}^{2}), & 0<\lambda<J<\lambda_{*}.
\end{array}\right.
\end{equation}
Since the ferromagnetic product state $\ket{\uparrow\uparrow\cdots\uparrow}$
is the ground state $\ket{g(\lambda_{*})}$ by taking $\lambda_{*}\to+\infty$,
then we obtain 
\begin{equation}
\lim_{t,N\to\infty}\frac{I(t)}{Nt^{2}}=\left\{ \begin{array}{cc}
\frac{J^{2}(4\lambda^{2}-3J^{2})}{\lambda^{4}}, & J<\lambda,\\
1, & 0<\lambda<J.
\end{array}\right.
\end{equation}

\end{document}